\documentclass[12pt, draftclsnofoot, onecolumn]{IEEEtran}
\usepackage{cite}
\usepackage{amsmath,amssymb,amsfonts}
\usepackage{algorithmic}
\usepackage{graphicx}
\usepackage{textcomp}
\usepackage{xcolor}
\usepackage{hyperref}

\newcommand{\NF}{$N_\mathrm{Encoder}$}
\newcommand{\Nf}{$N_\mathrm{Decoder}$}

\ifCLASSINFOpdf
\else
\fi
\ifCLASSOPTIONcompsoc
  \usepackage[caption=false,font=normalsize,labelfont=sf,textfont=sf]{subfig}
\else
  \usepackage[caption=false,font=footnotesize]{subfig}
\fi
\begin{document}
%
\title{Channelformer: Attention based Neural Solution for Wireless Channel Estimation and Effective Online Training}

%

\author{\IEEEauthorblockN{Dianxin Luan,~\IEEEmembership{Student Member,~IEEE,} John Thompson,~\IEEEmembership{Fellow,~IEEE}}\\
\IEEEauthorblockA{\textit{Institute for Digital Communications, School of Engineering, University of Edinburgh, Edinburgh, EH9 3JL, UK}\\
Email address : Dianxin.Luan@ed.ac.uk, john.thompson@ed.ac.uk}
}

\maketitle

\begin{abstract}
In this paper, we propose an encoder-decoder neural architecture (called Channelformer) to achieve improved channel estimation for orthogonal frequency-division multiplexing (OFDM) waveforms in downlink scenarios. The self-attention mechanism is employed to achieve input precoding for the input features before processing them in the decoder. In particular, we implement multi-head attention in the encoder and a residual convolutional neural architecture as the decoder, respectively. We also employ a customized weight-level pruning to slim the trained neural network with a fine-tuning process, which reduces the computational complexity significantly to realize a low complexity and low latency solution. This enables reductions of up to 70\% in the parameters, while maintaining an almost identical performance compared with the complete Channelformer. We also propose an effective online training method based on the fifth generation (5G) new radio (NR) configuration for the modern communication systems, which only needs the available information at the receiver for online training. Using industrial standard channel models, the simulations of attention-based solutions show superior estimation performance compared with other candidate neural network methods for channel estimation. 
\end{abstract}

\IEEEpeerreviewmaketitle

\section{Introduction}
\label{Introduction}
\IEEEPARstart{F}{or} fifth generation (5G) wireless communication systems and beyond, the orthogonal frequency division multiplexing (OFDM) baseband waveform will be retained \cite{dang2020should}, which requires precise channel state information in order to compensate for the channel distortion and provide robust communication \cite{shafi20175g}. Conventional channel estimation methods are the least-squares (LS) and minimum mean squared error (MMSE) approaches \cite{edfors1998ofdm}. However, with the development of modern communication systems, the LS method cannot achieve precise estimation and the implementation of the MMSE method is challenging as the perfect and complete channel statistics cannot be accessed accurately in advance. Moreover, conventional channel estimation solutions \cite{van1995channel} \cite{lottici2002channel} also cannot achieve sufficient performance. 

Meanwhile, artificial intelligence is impacting on the optimization and configuration of 6G \cite{letaief2019roadmap}. It motivates the researchers in the field of wireless channel estimation to explore neural network solutions for improved performance \cite{he2018deep} \cite{soltani2019deep} \cite{li2019deep} \cite{luan2021low}. Compared with the conventional methods which aim to find the closed-form expression, neural network methods are typical data-driven methods aiming for a satisfactory and local optimum solution. ChannelNet \cite{soltani2019deep}, ReEsNet \cite{li2019deep} and Interpolation-ResNet \cite{luan2021low} are recently published neural networks for wireless channel estimation. The success of the attention mechanism has been demonstrated in recent years \cite{yan2019stat}. Attention-based solutions have significant advantages over conventional neural networks, particularly with the release of the transformer \cite{vaswani2017attention}. BERT \cite{devlin2018bert} is a transformer-based pre-trained model for language processing and has proved to be efficient and effective. Compared with the convolutional neural networks, the vision transformer \cite{dosovitskiy2020image} also has superior performance. The attention mechanism is also investigated to improve the performance for wireless channel estimation in \cite{pan2021channel} \cite{lu2021channel} \cite{jiang2021dual}. The paper \cite{chen2020channel} exploits the transformer for the improved channel estimation. Reference \cite{mashhadi2021pruning} introduces non-local attention \cite{wang2018non} to achieve the channel estimation for MIMO-OFDM system. A graph attention network is utilized in \cite{tekbiyik2021channel} for the reconfigurable intelligent surface-assisted communications. An encoder-decoder architecture is also proposed in \cite{luan2022attention} by deploying the attention mechanism and the papers \cite{yang2021deep} \cite{gao2021attention} also use the attention mechanism. However, for the practical communication system, neural network channel estimation faces severe problems when implemented in the real world. 
\subsection{The challenges for neural network channel estimation}
\subsubsection{Input precoding}
For conventional channel estimation neural networks, LS is widely exploited as the input of the neural networks and each element of the input features is directly processed without any pre-processing. However, due to the fact that the correlation matrix of channel gains at different subcarriers may not be just diagonal, the importance of each channel gain is not necessarily equal for all the subcarriers. It indicates that some elements could have a strong correlation to the channel prediction while others are not very correlated, which means that the importance of the input features will differ. By allowing that some elements of the LS estimate could be more significant than others in the practical channels, the neural networks need to focus on a critical subset of the LS estimates. This is in contrast to most state-of-the art networks which process the uncorrelated elements equally with those key elements, because the low-correlated features can be considered as a noise in the input. Therefore, neural networks should pay attention to fewer but more important features. This paper exploits the property of the correlation by pre-processing the input features (called input precoding), aiming to improve the estimate performance. The MMSE method denoises the LS estimate with the correlation matrix assisting to improve the performance significantly. 
\begin{figure}[htbp]
\centerline{\includegraphics[width=0.75\textwidth]{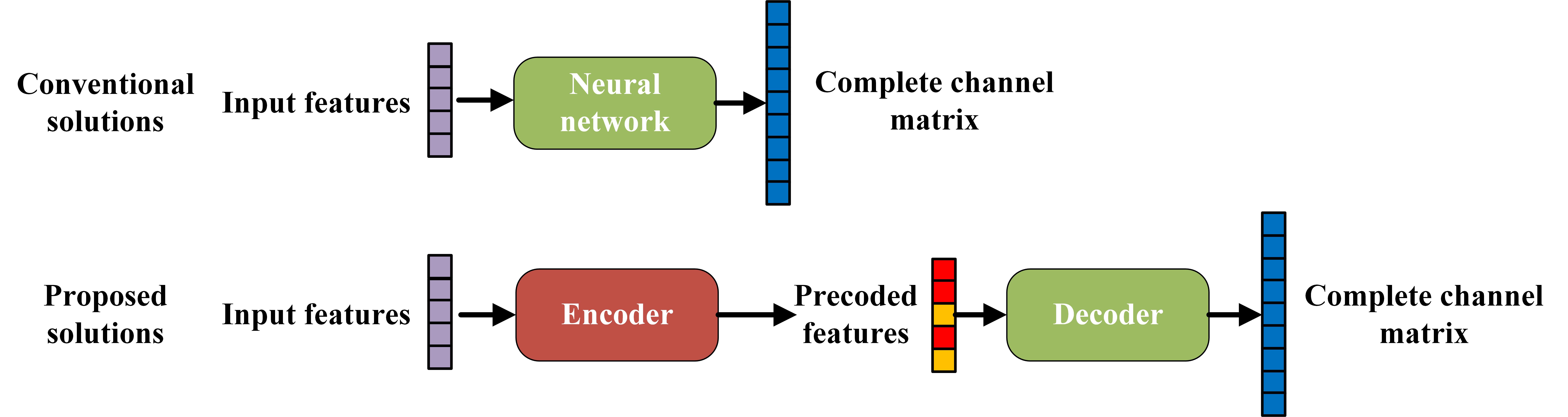}}
\caption{Comparison of the conventional neural network solution for channel estimation (top) and the proposed neural network (bottom).}
\end{figure}
\subsubsection{Online implementation}
The practical online implementation of neural networks should be capable of generalizing to extended conditions, supporting the trained neural networks to maintain consistent performance in the practical channels like COST 2100 channel \cite{liu2012cost} or the 3GPP TS38.901 fast fading channel. When the channel is found to be changed significantly from the channel for training and the measured performance is not acceptable, online training \cite{zheng2021online} \cite{jha2021online} \cite{yang2019deep} \cite{mei2021low} \cite{huang2018deep} is needed to fine-tune the neural network with online samples, to compensate for this degradation. However, it is a critical challenge to realize the online implementation for conventional neural networks in the real-world communication system. 
\begin{itemize}
    \item Realizability of online training: While online training is essential for practical implementation, the training process for many neural network solutions is not feasible in the real scenarios \cite{soltani2019deep} because a complete, noise-free channel matrix is not available. Assuming that the input of the neural network is also LS estimate, neural network solutions achieve both the frequency interpolation and time interpolation to predict the complete channel matrix commonly which can improve the performance to MMSE level or beyond, because these neural networks have already been trained with the perfect and complete channel information. When the channel is found to be changed significantly, the corresponding online training may require the perfect channel matrix of the complete data packet for training. The critical challenge is that this channel knowledge is unavailable from the real-world noisy receiver waveform. 
    \item Reliability of online training: The online training process needs to be reliable to achieve superior performance compared with conventional methods. However, the physical channel can be non-stationary when collecting the training samples, while training often needs sufficient samples to estimate the channel parameters properly. If the channel varies significantly among the collected training samples severely, especially in a single mini-batch, the online training process will randomly update the parameters. To avoid that, the online training dataset should be sampled within a controlled period to ensure the channel's coherence. Moreover, the training process should be robust to outlier samples. 
\end{itemize}
\subsubsection{Complexity and latency of implementing the neural network}
The complexity of the practical neural network solution should be considered carefully for low-latency and low-power communication systems because 5G often requires low-latency decoding. As the operation of multiplying matrices can be accelerated by at least 10 times \cite{blalock2021multiplying}, this work only considers the reduction of the network parameters and the critical path of the neural network to minimize the floating-point operations and the sequential latency. The latency under consideration involves the forward delay and online training latency. The forward delay refers to the running time for channel estimation, which depends on the complexity and architecture of the neural network. Online training delay refers to the latency of the online training process. Commonly, the training duration is considerable \cite{2021channel} \cite{ershadh2021computationally} \cite{jiang2021ai}, which means that the designed neural network solutions should have the property of fast convergence. Moreover, by using online nuclear norm-based loss function, the online training solution proposed in \cite{zheng2021online} calculates noisy measurements of a linear mapping for the unseen channels, which provides the channel information for neural networks to learn. It indicates that online training introduces an additional computation for the label generation and training. 
\subsection{Main contributions and outline}
To resolve the challenges mentioned, an encoder-decoder architecture called Channelformer is proposed for wireless channel estimation in the downlink scenario. Channelformer uses the multi-head attention and a convolutional neural network as the encoder and a residual convolutional architecture for the decoder. It achieves an improved performance to state-of-the-art neural network solutions and processing complexity is reduced. Our proposed method should perform robustly when dealing with some mismatches between the training and test datasets. However, the neural network needs to be retrained for significant changes, e.g. the number of subcarriers used. The major features of the proposed solution are 
\begin{itemize}
    \item To achieve the input precoding, we use an encoder involving the self-attention attention to pre-process the input features. The multi-head attention module helps capture the valuable information/features (the elements of LS estimate with a strong correlation to the channel predictions) and is expected to focus on different parts in different heads. Unlike conventional neural networks that process all the LS estimates equally, a pre-processing architecture is applied to focus on the most critical elements among the LS estimate. This will ensure that the subsequent neural network can focus on the critical features to yield the best channel estimate. Compared to HA02 \cite{luan2022attention}, the neural architecture of the encoder is modified for the Channelformer to provide both reduced complexity and improved performance. 
    \item We propose a simple and effective online training algorithm which can directly exploit the received signals for online training. The proposed training approach uses reference symbol information to generate the regression label. 
    The mini-batch size of the online training process is reduced to avoid random evolution of the Channelformer weights, which also simplifies the online training process and reduces the online training latency. Huber loss \cite{huber1965robust} is employed as loss function to balance the resistance to the outlier training samples and the convergence speed. Moreover, the encoder (multi-head attention module) is believed to accelerate the training processing because it captures the critical elements of features to focus on. We also investigate the generalization capability of the Channelformer to extended SNR and Doppler shift cases and provide more robust performance. 
    \item In order to compress the neural network and reduce latency, this paper customizes a weight-level pruning methodology for both the encoder and the decoder individually and minimizes the layers in the critical path. The subsequent fine-tuning training process also uses a smaller mini-batch size. Compared with the complete neural networks, the compressed neural networks retain almost identical performance and generalization capabilities. 
\end{itemize}

The rest of the paper is organized as follows. Section.~\ref{System architecture and channel deployed} presents the description of the baseband, frame structure based on the 5G New Radio (NR) standardization and the propagation channels. Section.~\ref{Conventional methods and neural networks for channel estimation} introduces the conventional methods LS and MMSE for channel estimation and describes the Channelnet, ReEsNet and Interpolation-ResNet networks for performance comparison. Section.~\ref{Channelformer_section} proposes the system description of Channelformer and the online training algorithm. Section.~\ref{Acceleration of the neural network} discusses the compression of the neural network to reduce the running latency and memory. Section.~\ref{Simulation results} compares the simulation results of the individual performance. Finally, Section.~\ref{Conclusion} summarizes the key findings of the paper. 
\section{System architecture and channel models}
\label{System architecture and channel deployed}
\subsection{OFDM baseband architecture}
\label{Baseband Architecture}
At the transmitter, the bit level signal $s(i)$ is processed in a Quadrature Phase Shift Keying (QPSK) modulator with Gray coding, and then pilot signals are inserted. Each slot consists of $N_s = 14$ OFDM symbols and each OFDM symbol involves $N_f = 72$ subcarriers, matching 6 5G NR resource blocks. The time-domain allocations for the demodulation reference signal (DM-RS) comprises 2 single symbol DM-RS waveforms and the 1\textsuperscript{st} and 13\textsuperscript{th} OFDM symbols are reserved for pilots ($N_{pilot} = 2$) \cite{dahlman20205g}. For the first pilot OFDM symbol, the indices of the pilot subcarriers start from the first subcarrier and are spaced by $L_s = 2$ subcarriers. For the second pilot symbol, the pilot subcarriers start at the second subcarrier and are also spaced by $L_s = 2$ subcarriers. The remaining subcarriers in the pilot symbols are set to 0. All of the data subcarriers in the remaining 12 OFDM symbols are assigned by QPSK-modulated symbols. The inverse fast Fourier transform (IFFT) converts the frequency domain data symbols to the time domain OFDM signal samples where $T_{Sym}$ denotes the duration of each pilot-data symbol. Then the cyclic prefix (CP) of duration $T_{CP}$ is added to the front of each symbol to provide resistance to multipath effects. Normally the fast Fourier transform (FFT) and IFFT operators use scaling factors of 1 and $1/{N_f}$, but those are changed to $1/\sqrt{N_f}$ to avoid changing the power of the FFT/IFFT outputs compared to the inputs. The channel is assumed to be a multipath fading channel with $M$ resolvable paths. By removing CP and converting the received data into the frequency domain by FFT operation, the received signal $\mathbf{Y} \in \mathbb{C}^{{N_f}\times N_{s}}$, can be represented by 
\begin{equation}
\mathbf{Y} = \mathbf{H} \circ \mathbf{X} + \mathbf{W} \\ , 
\end{equation}
Where $\mathbf{X},\ \mathbf{W}$ and $\mathbf{H} \in \mathbb{C}^{{N_f}\times N_{s}}$ are the Discrete Fourier Transforms (DFT) of the transmitted signal, the additive white Gaussian noise and the channel impulse responses. The operator $\circ$ represents the Hadamard product. With the condition that the maximum delay is less than the duration of CP, the received pilot signal is extracted to provide a channel amplitude and phase reference for each data symbol in the complete packet. Then the recovered data symbols are processed in the QPSK demodulator to obtain the received bit-level data estimates $\hat{s}(i)$. 
\subsection{Channel model and frame structure}
\label{Channel model and frame structure}
We consider a single-input-single-output (SISO) downlink scenario. The deployed channel is modelled as a Rayleigh fading channel by using the generalized method of exact Doppler spread method \cite{patzold2009two}. The power delay profiles (PDP) are the extended pedestrian A model (EPA), extended vehicular A model (EVA) and extended typical urban model (ETU) defined in 3GPP TS 36.101, representing a low, medium, and high delay spread environment respectively. It is assumed that the operating frequency is 2.1GHz (sub-6GHz band) and the subcarrier spacing is $15$kHz. Moreover, the sampling rate is $1.08$MHz and $T_{CP}$ is 16 samples. 
\section{Conventional methods and neural networks for channel estimation}
\label{Conventional methods and neural networks for channel estimation}
The conventional channel estimation methods are the least squares (LS), Decision-directed channel estimation (DD-CE) and minimum mean squared error (MMSE) used to compare performance with neural network solutions. The algorithm implementations of the conventional methods are explained, and the deployed neural network methods are also introduced. 
\subsection{LS method}
\label{LS method}
By minimizing the mean squared error (MSE) between $\mathbf{Y}$ and $\mathbf{H} \circ \mathbf{X}$ at the pilot positions, i.e. $\mathop{\arg\!\min_{\mathbf{H}}}\Arrowvert \mathbf{Y}-\mathbf{H} \circ \mathbf{X}\Arrowvert^{2}_{2}$ at the pilot positions, to give an estimate of $\mathbf{H}$, the frequency domain LS estimation is given by 
\begin{equation}
    \mathbf{\hat{H}_{LS} = \frac{Y_{Pilot}}{X_{Pilot}}} , 
\end{equation}
Where $\mathbf{Y_{Pilot}}, \mathbf{X_{Pilot}} \in \mathbb{C}^{\frac{N_f}{2}\times N_{pilot}}$ denotes the received and transmitted pilot signals respectively, and the mathematical division operation is performed element-wise. LS is easy to implement with extremely low complexity for all the existing channels. When we evaluate the MSE of LS method for each pilot OFDM symbol, MSE is inversely proportional to the numerical value of SNR from equ.~(\ref{LS}), 
\begin{equation}
\label{LS}
    \mathrm{MSE} = E\left\{\left(\mathbf{H_{ref}} - \mathbf{\hat{H}_{LS}}\right)^{H}\left(\mathbf{H_{ref}} - \mathbf{\hat{H}_{LS}}\right)\right\} = \frac{\sigma_N^2}{\sigma_X^2} , 
\end{equation}
Where $\mathbf{H_{ref}}$ is the noise-free channel vector at the LS-corresponding position and $\frac{\sigma_X^2}{\sigma_N^2}$ denotes the numerical value of SNR. The prediction $\mathbf{\hat{H}_{LS}}$ is then resized by bilinear interpolation \cite{press1989numerical} in both the time and frequency domain to estimate the complete channel matrix $\mathbf{\hat{H}_{LS}^{frame}} \in \mathbb{C}^{N_f \times N_s}$. 
\subsection{DD-CE method}
To provide an alternative non-AI method for comparison, we implemented an iterative method DD-CE \cite{liu2014channel}. The $\hat{H}_{LS}^{frame}$ is used to detect the QPSK data symbols using the equation $\mathbf{\hat{X}} = \mathbf{Y/\hat{H}_{LS}^{frame}}$ and then hard decisions are applied to all entries of $\mathbf{\hat{X}}$ to map to the nearest QPSK constellation point. These QPSK symbols are then fed back to update the LS estimate $\mathbf{\hat{H}_{LS}^{frame} = Y/\hat{X}}$. The iterations continue until the hard decisions do not change or one hundred iterations are reached. The final DD-CE prediction is denoised in the frequency domain by the Wiener filter used in \cite{edfors1998ofdm}. However, the DD-CE method still suffers from error propagation under high Doppler frequency or low SNR conditions. 
\subsection{FD-MMSE and 2D-MMSE methods}
\label{MMSE method}
To minimize the square of Euclidean distance between the actual channel matrix and the LS estimate for each pilot symbol, the one-dimensional (1D) frequency domain linear MMSE channel estimation of the $i^{th}$ OFDM symbol \cite{omar2008performance} $\mathbf{\hat{H}_{MMSE}} \in \mathbb{C}^{N_f \times N_{pilot}}$ is computed by 
\begin{equation}
\label{MMSE}
    \mathbf{\hat{H}_{MMSE}}(j) = \mathbf{R_{hh_{p}}}(j)\left(\mathbf{R_{h_{p}h_{p}}}(j) + \frac{\sigma_N^2}{\sigma_X^2}I\right)^{-1}\mathbf{\hat{H}_{LS}}(j), \ \mathrm{with} \ j = \begin{cases}
    1, & i=1 \\
    2, & i=13 \\
\end{cases}, 
\end{equation}
Where $j$ denotes the index of the pilot symbol, $\mathbf{h}(i) \in \mathbb{C}^{N_f}$ is the channel gain vector for the $i^{th}$ OFDM symbol and $\mathbf{h_p}(i) \in \mathbb{C}^{\frac{N_f}{2}}$ denotes the channel gain matrix for the $i^{th}$ OFDM symbol which contains $\frac{N_f}{2}$ pilot subcarriers. $I$ is the corresponding identity matrix. Assuming the noise-free channel vectors at the pilot symbol locations are known, the correlation matrices are computed by 
\begin{equation}
\begin{split}
    \mathbf{R_{hh_p}}(j) &= E\left\{\mathbf{h(j)h_{p}(j)^{H}}\right\}, \\ 
    \mathbf{R_{h_ph_p}}(j) &= E\left\{\mathbf{h_{p}(j)h_{p}(j)^{H}}\right\}. 
\end{split}
\label{correlation}
\end{equation}
To predict the channel matrix for the whole slot, linear interpolation \cite{press1989numerical} is also implemented to achieve time interpolation for this 1D FD-MMSE method. 

To achieve the two-dimensional (2D) FD-MMSE method, the conventional non-AI methods are challenging because despite the accuracy, statistical models are cumbersome and difficult to handle \cite{tang2007pilot} \cite{nissel2018doubly}. Therefore, many papers are based on a basis expansion model (BEM) \cite{tang2007pilot} \cite{giannakis1998basis} \cite{borah1999frequency} for the doubly-selective channels. To interpolate the 1D FD-MMSE method in the time domain precisely, the correlation matrix at each data symbol is calculated by equ.~(\ref{correlation}) by exploiting each noise-free channel vector at the data OFDM symbols and the corresponding LS estimates at each OFDM symbol are also assumed to be known. With the perfect and complete channel knowledge known, the 2D FD-MMSE method is the most precise method to compute the channel matrix, which should give the best performance for methods that use the mean squared error loss function. By utilizing prior statistical knowledge of the channel state, the MMSE method improves the performance of the LS method but requires unavailable channel information. Therefore, physical implementation of the MMSE method is difficult which motivates us to research neural network solutions. 
\subsection{ChannelNet, ReEsNet, TR and Interpolation-ResNet}
Due to the considerable performance gap between the LS method and the MMSE method, the neural network solutions are investigated for the improvement. ChannelNet \cite{soltani2019deep} (670,000 parameters) is one of the first released neural network solutions for channel estimation. However, conventional deep neural networks may not exhibit the potential degeneracy, which inspires the researchers to utilize a residual architecture to solve this. ReEsNet \cite{li2019deep} is a residual convolutional neural network with 53,000 parameters, which outperforms ChannelNet. Compared with ReEsNet, Interpolation-ResNet with only 9,442 parameters (called \textsl{InterpolateNet} in this paper) \cite{luan2021low} achieves a slightly improved performance and 82\% reduced parameters. However, the generalization capability of InterpolateNet and ReEsNet trained \cite{luan2021low} is quite limited, which motivates us to research with the attention mechanism to propose HA02 \cite{luan2022attention}. As the other neural network solutions are much less complex than ChannelNet, we only consider the state-of-the-art networks when presenting simulation results. TR \cite{chen2020channel} is a transformer-based solution which involves a transformer encoder \cite{vaswani2017attention} and a 1D-CNN. The output is resized by bilinear interpolation to estimate the channel matrix of the whole slot. 
\subsection{Discussion on the performance of neural networks}
\label{Discussion on the performance of neural network}
Deep learning based channel estimators, which exploit the LS estimate as input, denoise and interpolate the input in both the frequency and time dimension to approach the actual channel matrix. To minimize the distance between the LS estimate and the actual channel matrix, the neural networks are believed to find a local optimization while the MMSE method with perfect channel knowledge is close to global optimization. Therefore, the 1D FD-MMSE method should outperform the neural network solutions on denoising and frequency interpolation. However, accurate time interpolation provides the possibility to outperform the 1D FD-MMSE method because it uses linear interpolation in the time domain. By exploiting the perfect channel matrix of the complete package for training, the neural networks can outperform the 1D FD-MMSE method even though it is not a practical solution. Therefore, we propose two versions of the proposed method (offline Channelformer and online Channelformer). For the offline Channelformer, the prediction is the channel matrix of whole slot. For the online Channelformer, it is only capable of denoising and frequency interpolation. In this paper, a united interpolation method for LS, 1D FD-MMSE and online Channelformer is used to achieve time interpolation for estimation. Therefore, the online Channelformer cannot outperform the 1D FD-MMSE method because both methods exploit same channel information. 
\section{Channelformer}
\label{Channelformer_section}
HA02 \cite{luan2022attention} is an encoder-decoder neural network solution that uses the transformer encoder \cite{vaswani2017attention} to improve performance. However, the computational complexity can be reduced for the low-complexity applications and the performance can be improved further, so this paper shows how this can be achieved. The self-attention mechanism is also implemented to focus on the features that are \textbf{small but important}. The motivation of implementing the self-attention mechanism is to achieve the input precoding by exploring the significance of the feature's elements, instead of processing all the feature's elements as being equally important. 

\begin{figure}[htbp]
\centerline{\includegraphics[width=0.9\textwidth]{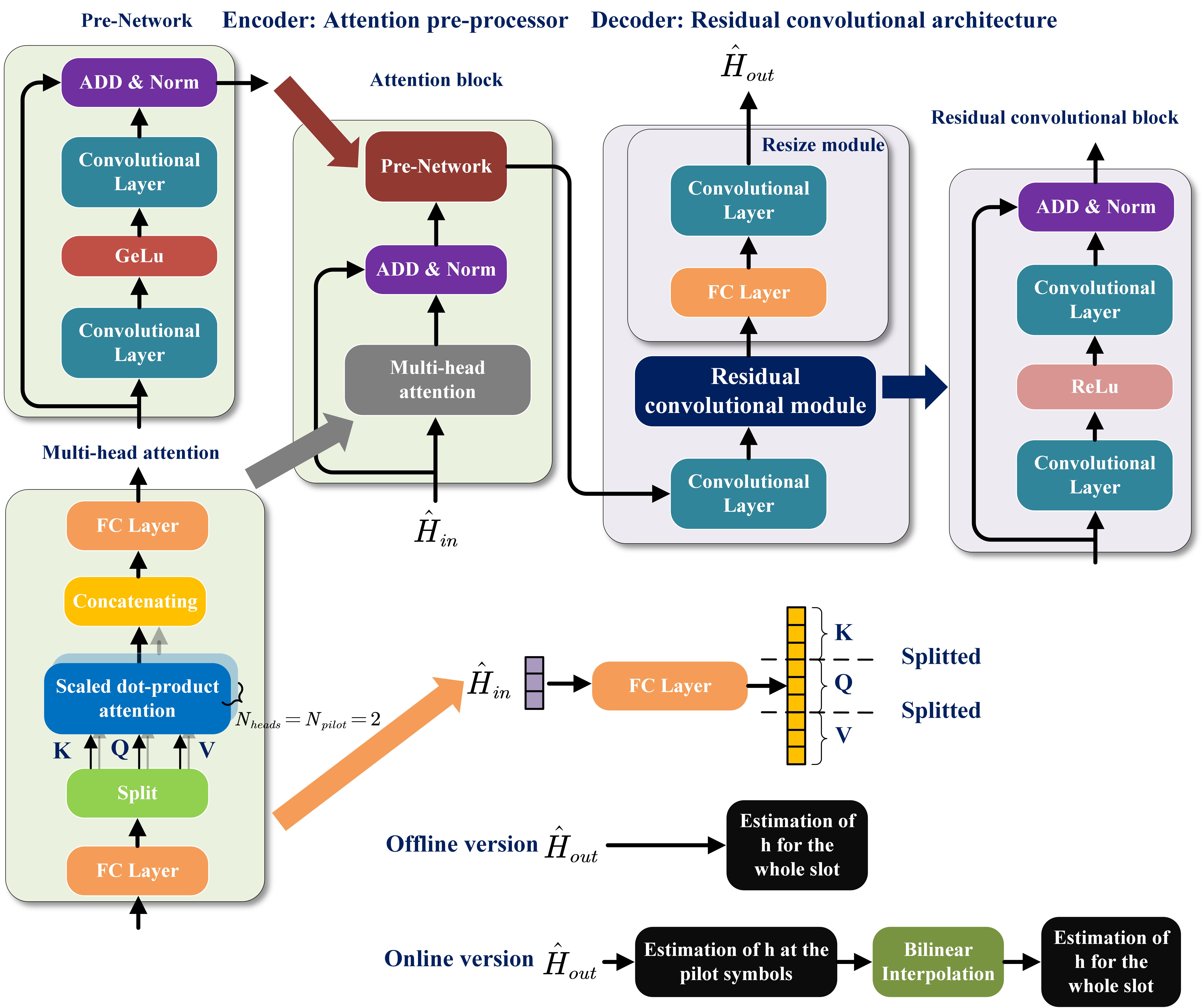}}
\caption{A detailed overview of the Channelformer neural network architecture. The bottom right of the figure shows the offline version (top) and the online version (bottom).}
\label{Channelformer}
\end{figure}

For the input of Channelformer, the matrix $\mathbf{\hat{H}_{LS}}$ is concatenated to be one column vector $\in \mathbb{C}^{\frac{N_{pilot}N_f}{L_s}}$ and the real and imaginary parts of the one column vector are split into two different channels as the second dimension of the input array. Therefore, the input is the transformed estimation of LS $\mathbf{\hat{H}_{in}}$ has a size of $\mathbb{R}^{\left(\frac{N_{pilot}N_f}{L_s}\right) \times 2}$. The offline Channelformer is trained by using the channel matrix of the whole slot and the corresponding complex estimate of Channelformer is denoted as $\mathbf{\hat{H}_{Channelformer}} \in \mathbb{C}^{N_f \times N_{s}}$. The online Channelformer is trained by using the channel matrix at the position of the pilot symbols instead of the channel matrix of the whole slot. It is compatible with the realistic online training solution proposed. Therefore, the output of online Channelformer $\mathbf{\hat{H}_{out}}$ has a size of $\mathbb{R}^{(N_{pilot}{N_f}) \times 2}$ where the index 1 of the second dimension is the real part and index 2 is the imaginary part. The first $N_f$ samples in each dimension are the prediction for the first pilot symbol and the second set of $N_f$ samples are for the second pilot symbol. Therefore, the corresponding complex estimate, denoted as $\mathbf{\hat{H}_{Channelformer}}$, has a size of $\mathbb{C}^{N_f \times N_{pilot}}$. Channelformer shown in Fig.~\ref{Channelformer} involves two substructures, which are attention pre-processor and residual convolutional architecture. 
\subsection{Encoder: Attention pre-processor}
\label{Encoder: Attention stack}
In this paper, we implement the multi-head attention following \cite{vaswani2017attention}. The scaled dot-product attention module has three inputs named the query ($\mathbf{Q}$), key ($\mathbf{K}$) and value ($\mathbf{V}$) generated from the same input after a linear transformation realized by fully-connected layer. By mapping the query $\mathbf{Q}$ and key $\mathbf{K}$ to generate the probability of the $\mathbf{V}$, the expectation of $\mathbf{V}$ is calculated by multiplying the computed probability with $\mathbf{V}$'s corresponding value, where input precoding is achieved. The encoder is composed of one attention block. 
\subsubsection{\textbf{Multi-head attention and normalization}}
\label{Multi-head attention and normalization}
The first fully-connected layer resizes the input $\mathbf{x} \in \mathbb{R}^{\left(\frac{N_{pilot}N_f}{L_s}\right) \times 2}$ to $\mathbf{y} \in \mathbb{R}^{\left(\frac{3N_{pilot}N_f}{L_s}\right) \times 2}$ by 
\begin{equation}
    \mathbf{y} = \mathbf{Wx} + \mathbf{b}, 
\label{fc}
\end{equation}
where $\mathbf{W} \in \mathbb{R}^{\left(\frac{3N_{pilot}N_f}{L_s}\right) \times \left(\frac{N_{pilot}N_f}{L_s}\right)}$ is the weight, $\mathbf{b} \in \mathbb{R}^{\left(\frac{3N_{pilot}N_f}{L_s}\right) \times 1}$ is the bias. Multi-head attention splits that output equally to 3 sub-inputs as the $\mathbf{K}$, $\mathbf{Q}$ and $\mathbf{V}$ $\in \mathbb{R}^{\left(\frac{N_f}{L_s}\right) \times 2 \times N_{head}}$, for $N_{heads}$ heads with the constraint that $N_{heads}$ = $N_{pilot}$. In different heads, the attention mechanism focus on different features. Moreover, we do not apply causal mask and dropout in the multi-head attention block \cite{vaswani2017attention} because we do not need both. For the i\textsuperscript{th} element of the feature, it can be correlated to the non-causal elements (position larger than i) from the correlation matrix of channels, which indicates that the features may not be sequential. Moreover, the channel noise can realize regularization. Therefore, we remove the dropout and the L2 regularization is set to a value close to $0$. The scaled dot-product attention $\in \mathbb{R}^{\left(\frac{N_f}{L_s}\right) \times 2 \times N_{heads}}$ \cite{vaswani2017attention} is computed by 
\begin{equation}
    \mathbf{Attention} = \mathrm{softmax}\left(\frac{\mathbf{QK}^{T}}{\sqrt{d_{k}}}\right)\mathbf{V}, 
\label{scaled dot-product attention}
\end{equation}
Where $d_{k} = \frac{N_f}{L_s}$ is a scaling factor to normalize $\mathbf{QK}^{T}$. After concatenating the outputs from scaled dot-product attention for different heads in the first dimension, the concatenated result $\in \mathbb{R}^{\left(\frac{{N_f}{N_{heads}}}{L_s}\right) \times 2}$ is forwarded to the second fully-connected layer to generate the output of the attention module $\in \mathbb{R}^{\left(\frac{{N_f}{N_{heads}}}{L_s}\right) \times 2}$. The Add \& Norm layer adds the linear transformed output from the scaled dot-product attention and the skip connection from the first fully-connected layer input, and applies layer normalization \cite{ba2016layer} to the superimposed result $\in \mathbb{R}^{\left(\frac{{N_f}{N_{pilot}}}{L_s}\right) \times 2}$. The operation of layer normalization is given in equ.~(\ref{layer}). For each head's input $\mathbf{x} \in \mathbb{R}^{\frac{N_{pilot}N_f}{L_s} \times 1}$, a subset of the superimposed result that only consists of the whole elements of the superimposed result's first dimension, the layer normalization computes the corresponding output $\mathbf{y}$ by 
\begin{gather}
    \mathbf{y} = \mathbf{w}\left(\frac{\mathbf{x} - \mu}{\sqrt{\sigma^{2} + \varepsilon}}\right) + \mathbf{b} , \\
    \mu = \mathrm{mean}(\mathbf{x}) = \left(\frac{L_s}{N_{pilot}N_f}\sum_{i = 1}^{\frac{N_{pilot}N_f}{L_s}} x_{i}\right) , \\
    \sigma^{2} = \mathrm{mean}\left(\left\Arrowvert\mathbf{x} - \mu\right\Arrowvert_{2}^{2}\right) , 
\label{layer}
\end{gather}
Where $\varepsilon = 10^{-5}$, $\mu$ denotes the empirical mean of $\mathbf{x}$ and $\sigma^{2}$ is the corresponding variance. $\mathbf{w}, \ \mathbf{b} \in \frac{N_{pilot}N_f}{L_s} \times 1$ is the weight and bias which achieve the same linear mapping for each $\mathbf{x}$, which means that $\mathbf{w}$ and $\mathbf{b}$ is unchanged for each $\mathbf{x}$. The output of that Add \& Norm layer $\in \mathbb{R}^{\left(\frac{N_{pilot}N_f}{L_s}\right) \times 2}$ is then processed by the Pre-Network module. 
\subsubsection{\textbf{Pre-Network}}
It involves three layers, which are one convolutional layer with \NF \ filters kernel size of {2 $\times$ 2 $\times$ 1} with corresponding output $\in \mathbb{R}^{\left(\frac{N_{pilot}N_f}{L_s}\right) \times 2 \times N_{Encoder}}$, one activation layer and another convolutional layer with one filter kernel size of {2 $\times$ 2 $\times$ \NF} in series. The approximate GeLu \cite{hendrycks2016gaussian} is employed as the activation function for the encoder, which computes the output $\mathbf{y} \in \mathbb{R}^{\left(\frac{N_{pilot}N_f}{L_s}\right) \times 2 \times N_{Encoder}}$ from the input $\mathbf{x} \in \mathbb{R}^{\left(\frac{N_{pilot}N_f}{L_s}\right) \times 2 \times N_{Encoder}}$ by 
\begin{equation}
    y = 0.5x\left(1+\mathrm{tanh}\left(\sqrt{\frac{2}{\pi}}\left(x+0.044715x^{3}\right)\right)\right) . 
\label{GeLu}
\end{equation}
The Add \& Norm layer adds the output from the second convolutional layer $\in \mathbb{R}^{\left(\frac{N_{pilot}N_f}{L_s}\right) \times 2 \times 1}$ and the skip connection from the input of the Pre-Network $\in \mathbb{R}^{\left(\frac{N_{pilot}N_f}{L_s}\right) \times 2 \times 1}$, and applies the layer normalization \cite{ba2016layer} to the superimposed result in the first dimension. That normalized result is the input $\in \mathbb{R}^{\left(\frac{N_{pilot}N_f}{L_s}\right) \times 2 \times 1}$ to the decoder. Compared with HA02 \cite{luan2022attention}, the fully-connected layers are replaced by convolutional layers to reduce the number of parameters for the encoder and improve the performance. 
\subsection{Decoder: Residual convolutional architecture}
\label{Decoder: Residual convolutional architecture}
The encoded features $\in \mathbb{R}^{\left(\frac{N_{pilot}N_f}{L_s}\right) \times 2 \times 1}$ are then processed by the decoder, as shown in Fig.~\ref{Channelformer}. It is composed of (i) one convolutional layer with an output $\in \mathbb{R}^{\left(\frac{N_{pilot}N_f}{L_s}\right) \times 2 \times N_{Decoder}}$, (ii) one residual convolutional module which consists of K stacks of residual convolutional blocks and the output size is $\in \mathbb{R}^{\left(\frac{N_{pilot}N_f}{L_s}\right) \times 2 \times N_{Decoder}}$, and (iii) one upsampling module in series. The first convolutional layer has \Nf \ filters with kernel size of $\Gamma$. The residual convolutional block consists of one convolutional layer with \Nf \ filters (the corresponding output size is $\in \mathbb{R}^{\left(\frac{N_{pilot}N_f}{L_s}\right) \times 2 \times N_{Decoder}}$), followed by one ReLu layer with the corresponding output size of $\in \mathbb{R}^{\left(\frac{N_{pilot}N_f}{L_s}\right) \times 2 \times N_{Decoder}}$ and one convolutional layer with \Nf \ filters (the corresponding output size is $\in \mathbb{R}^{\left(\frac{N_{pilot}N_f}{L_s}\right) \times 2 \times N_{Decoder}}$). The kernel size of each is $\Gamma$. ReLu computes the output $\mathbf{y}$ from input $\mathbf{x}$ element-wise by 
\begin{equation}
    y = \begin{cases}
x, & x > 0 \\
0, & x \leq 0 \\
\end{cases} , 
\label{ReLu}
\end{equation}
Where $\mathbf{x}, \mathbf{y} \in \mathbb{R}^{\left(\frac{N_{pilot}N_f}{L_s}\right) \times 2 \times N_{Decoder}}$. The Add \& Norm layer processes the superimposed result $\in \mathbb{R}^{\left(\frac{N_{pilot}N_f}{L_s}\right) \times 2 \times N_{Decoder}}$ of the input and output $\in \mathbb{R}^{\left(\frac{N_{pilot}N_f}{L_s}\right) \times 2 \times N_{Decoder}}$ of the residual block to the upsampling module. The upsampling module consists of one fully-connected layer and one convolutional layer. Redundant parameters \cite{zhang2017defense} are found to be capable of resisting the unknown changes. However, we think redundant architecture, rather than redundant parameters, is the key for resistance and we conclude that in Section.~\ref{Complexity reduction: weight-level slimming and denoise gain}. Therefore, to achieve one-dimensional upsampling for the first dimension and improve the generalization to unknown SNR and Doppler shift by using redundant architecture, we deploy a fully-connected layer in the resize module. The last coupled convolutional layer has one filter with the kernel size of $\Gamma$ to generate the output $\mathbf{\hat{H}_{out}}$. 
\begin{itemize}

    \item For the offline Channelformer, we use $K$ = 3, \NF \ equal to 5, \Nf \ equal to 12, and $\Gamma$ = {5 $\times$ 5}. The fully-connected layer resizes the output of residual block from $\mathbb{R}^{\frac{N_{pilot}N_f}{L_s} \times 2 \times N_{Decoder}}$ to the size of $\mathbb{R}^{\left(N_{s}{N_f}\right) \times 2 \times N_{Decoder}}$. For each channel of the residual block's output $\mathbf{x_{D}} \in \mathbb{R}^{\frac{N_{pilot}N_f}{L_s} \times 2}$, this fully-connected layer resize each of them by equ.~\ref{fc} with the weight $\mathbf{w} \in \mathbb{R}^{\left(N_{s}N_{f}\right) \times \left(\frac{N_{pilot}N_f}{L_s}\right)}$ and the bias $\mathbf{b} \in \mathbb{R}^{\left(N_{s}N_{f}\right) \times 1}$. The corresponding channel estimation $\mathbf{\hat{H}_{Channelformer}} \in \mathbb{C}^{N_f \times N_{s}}$ is the estimate $\mathbf{H}$ of the whole slot. The total number of parameters is 117,659: 21,358 for the encoder and 96,301 for the decoder. Moreover, the major parameters are contributed by fully-connected layers. 
    
    \item For the online Channelformer, we use $K$ = 1, \NF \ equal to 5, \Nf \ equal to 2, and $\Gamma$ = {2 $\times$ 2}. The fully-connected layer resizes the output of residual block from $\mathbb{R}^{\frac{N_{pilot}N_f}{L_s} \times 2 \times N_{Decoder}}$ to the size of $\mathbb{R}^{\left(N_{pilot}{N_f}\right) \times 2 \times N_{Decoder}}$ in the same way with the offline Channelformer by the weight $\mathbf{w} \in \mathbb{R}^{\left(N_{pilot}N_{f}\right) \times \left(\frac{N_{pilot}N_f}{L_s}\right)}$ and the bias $\mathbf{b} \in \mathbb{R}^{\left(N_{pilot}N_{f}\right) \times 1}$. The corresponding channel estimation $\mathbf{\hat{H}_{Channelformer}} \in \mathbb{C}^{N_f \times N_{pilot}}$ is interpolated linearly \cite{press1989numerical} to predict the channel matrix of the whole slot, which fairly compares with the LS and FD-MMSE methods. The total number of parameters is 32,069: 21,358 for the encoder and 10,711 for the decoder. 98.34\% parameters are contributed by fully-connected layers, which indicates a further complexity reduction by weight-level pruning. For the configuration of online Channelformer, it is impossible to exceed the performance of FD-MMSE as discussed in Section.~\ref{Discussion on the performance of neural network}. 
    
\end{itemize}
\subsection{Online training algorithm}
\begin{figure}[htbp]
\centerline{\includegraphics[width=1\textwidth]{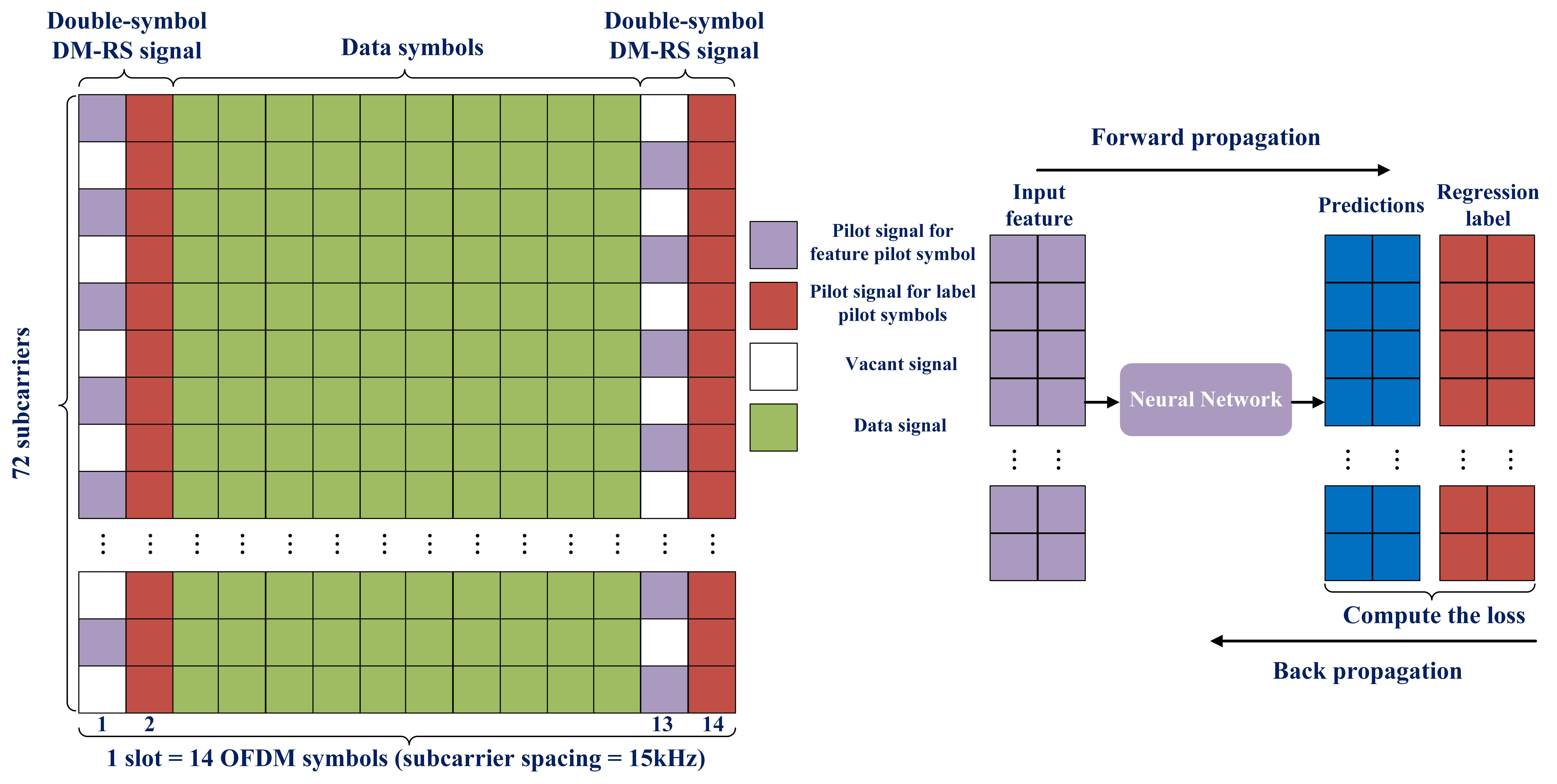}}
\caption{Frame description (double-symbol DM-RS model) on the left and online training process on the right.}
\label{Online training process}
\end{figure}

For the online deployment, online training is required to track the physical channels and achieve reliable communication. Instead of replacing the loss function, we propose an approach which yields a simple and effective online training algorithm. To improve the performance without access to the noise-free and complete channel matrix, the proposed online training process starts with the offline-trained neural network and exploits the denoised LS estimate $\mathbf{\hat{H}_{LS}}$ for the regression label, which is also synchronous with the transmission of the data. Rather than the single-symbol DM-RS for offline training described in Section.~\ref{Baseband Architecture}, the pilot pattern is changed to a non-standardized customized double-symbol DM-RS described in Fig.~\ref{Online training process} for the online training. For the label pilot symbols in Fig.~\ref{Online training process}, the pilot signal with higher SNR than the data signal takes all the subcarriers of that pilot symbol. This paper considers two different label pilot symbols design to achieve higher SNR for comparison. 
\begin{itemize}

    \item \textsl{\textbf{Power Increased Pilot Symbols:}} The power increased pilot label symbols contain data that is known to the receiver. A power increase of 5dB for the label pilots compared to the feature pilots is considered to guide the neural network to denoise by that 5dB gain, because the SNR of the label pilot symbol is 5dB higher than the feature pilot symbol for the adjacent pilot symbols of DM-RS. The average power of the whole slot is increased by 1.13dB, therefore, the power efficiency consumed is negligible to achieve the power-enhanced online training. 
    
    \item \textsl{\textbf{MMSE Processing of the label pilot Data:}} This method requires the receiver to process the received label pilot symbols by an estimated MMSE filter with the delay distribution given by equ.~(\ref{Density function}) to work. By contrast, the power of the label pilot symbols is the same as the data symbols, and an estimated MMSE method is exploited to denoise the label pilot symbols for training. The channel gains at the positions of the label pilot symbols are first estimated by the LS method described in Section.~\ref{LS method}, to provide the corresponding $\hat{H}_{LS}$ for the FD-MMSE method described in equ.~(\ref{MMSE}). To compute the correlation matrix required by equ.~(\ref{MMSE}), the $i,j^{th}$ element of the correlation matrix is computed by equ.~(\ref{label}) by letting $\tau_{rms} \to \infty$ and normalizing $r_{i, i} = 1, \ \forall{i \leq N_f}$. Each $\tau_{m}$ for the $M$ different paths is assumed to be independent. 
    \begin{equation}
    \begin{split}
        r_{i,j} &= \int ... \int \prod_{m=0}^{M-1}f_{\tau_{m}}\left(\tau_{m}\right)\left[\sum_{k=0}^{M-1} \theta\left(\tau_{k}\right)e^{-j2\pi\tau_{k}\left(i-j\right)/N_{f}}\right]d\tau_{0}...d\tau_{M-1} , \\ 
        &= \frac{1-e^{-2\pi jT_{CP}\left(i-j\right)/N_{f}}}{2\pi jT_{CP}\frac{i-j}{N_{f}}} , 
    \end{split}
    \label{label}
    \end{equation}
    Where $\theta\left(\tau\right) = \frac{T_{CP}e^{-\tau/\tau_{rms}}}{M\tau_{rms}\left(1-e^{-\frac{T_{CP}}{\tau_{rms}}}\right)}$ is assumed to be the exponentially decaying power delay profile and $f_{\tau}\left(\tau\right)$ is the probability density function of each $\tau_{k}$ given by \cite{edfors1998ofdm} 
    \begin{equation}
        f_{\tau}\left(\tau\right) = \begin{cases}
        1/T_{CP}, & \tau \in [0, T_{CP}] \\
        0, & otherwise \\
    \end{cases} . 
    \label{Density function}
    \end{equation}
\end{itemize}

The resulting FD-MMSE filter can be computed offline to minimize the online processing of the label data. The feature pilot symbol and the label pilot symbol of each double-symbol DM-RS are reserved for the input feature and the regression label of the feature pilot symbols. Moreover, the Doppler shift affects the performance of the proposed online training algorithm. The coherence time should exceed the duration of the adjacent double-symbol DM-RS signal significantly, to lessen the effects derived by the shift of the regression labels in the time domain. For a Rayleigh fading channel, the variation of the channel's correlation matrix caused by time shifts from $l$th OFDM symbol to $l'$th OFDM symbol can be calculated \cite{jakes1994microwave} by equ.~(\ref{estimated}) for the configuration described in Section.~\ref{Channel model and frame structure}. 
\begin{equation}
    r_t(l - l') = J_{0}\left(2\pi f_{Doppler}\left(T_{Sym}+T_{CP}\right)\left(l-l'\right)\right) , 
\label{estimated}
\end{equation}
Where $J_{0}()$ denotes the zeroth order Bessel function of the first kind. Even for a very high speed train scenario, where the Doppler frequencies are in the range 750-972 Hz, the corresponding $r_t$ varies from 0.94 to 0.97 for the adjacent OFDM symbols ($l - l' = 1$). This indicates that the transfer's effect on the online training process is not significant because the channel vectors at the feature pilot OFDM symbol and the label pilot symbol deployed in this paper are strongly-correlated. Moreover, the proposed double-symbol DM-RS is one example of this scheme and other pilot configurations ($l - l' \neq 1$) are possible. 
\section{Compression of the neural network}
\label{Acceleration of the neural network}
%
As the neural network should be compressed for online implementation to reduce the latency for low-power applications, we consider the compression of the neural network as both decreasing the critical path and reducing the computations of each layer \cite{rodgers1985improvements}, which then reduce the delay $\mathcal{T}$ of a forward pass from equ.~(\ref{Forward delay}). 
\begin{equation}
\mathcal{T} = \sum_{i = 1}^{N} \tau(i) , 
\label{Forward delay}
\end{equation}
Where $N$ represents the number of layers and $\tau(i)$ denotes the execution time of the $i^{th}$ layer. In this paper, we deploy the customized weight-level pruning to slim the trained neural network and reduce the computations of the connection layers with a high pruning ratio (70\% or more). Moreover, the number of layers is also considered to be minimized. 
\subsection{Reducing the critical path length: reduce number of layers}
The layer-level critical path of the neural network is reduced by decreasing the number of layers, as sequential delay cannot be parallelised because the next layer needs the output of the previous layer to proceed. Online Channelformer minimizes the layers by using one attention block at the encoder and one residual convolutional block at the decoder. 
\subsection{Slim the trained neural networks: reduce the execution time of each layer}
\begin{figure}[htbp]
\centerline{\includegraphics[width=0.6\textwidth]{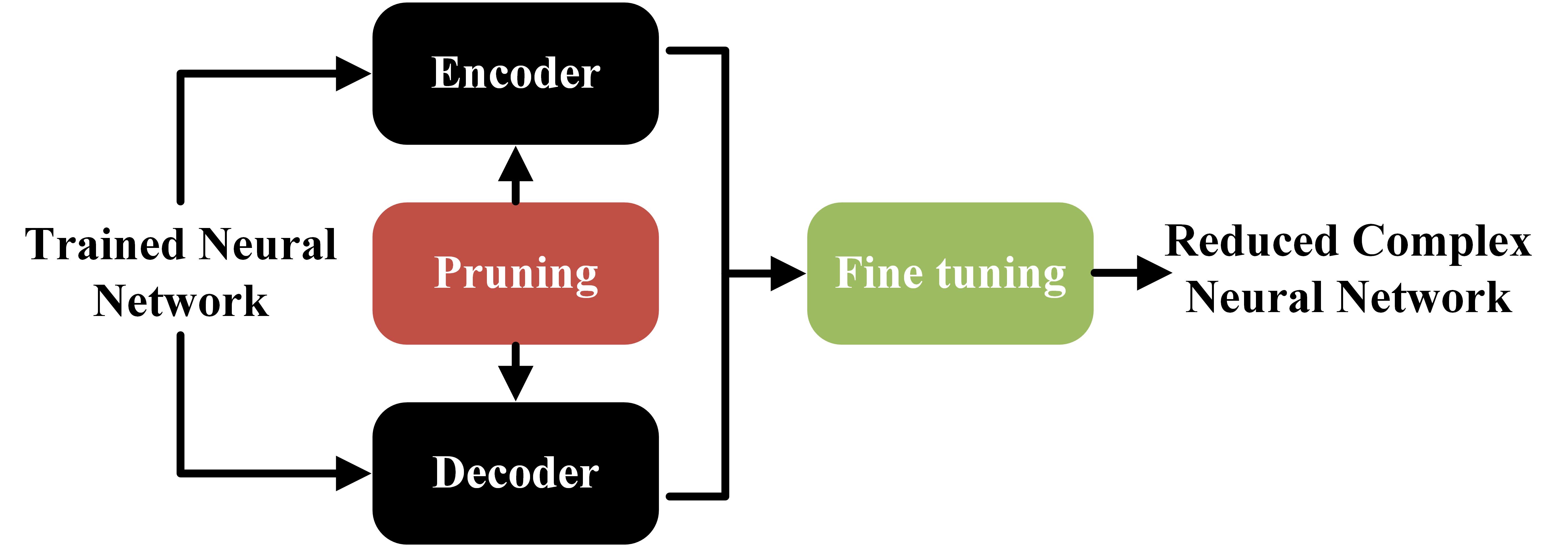}}
\caption{The customized weight-level pruning method used in this paper.}
\label{Weight-level slimming}
\end{figure}

To reduce the computations of the activation layers, GeLu is replaced by ReLu to lower the computations in the decoder as the computations compared by equ.~(\ref{GeLu}) and equ.~(\ref{ReLu}). For the computations of the connection layers, the optimization is achieved by applying the customized weight-level pruning in this paper. Weight-level pruning is the complexity reduction method that removes the redundant neural connections, and is widely discussed \cite{liu2017learning} \cite{takeda2019mimo} \cite{van2020optcomnet}. Removing a significant number of neurons degrades the performance, therefore, the pruning method often uses a fine-tuning process to compensate. The conventional pruning method \cite{han2015learning} is complicated for the online low-latency and low-power deployment. Therefore, we customize the weight-level pruning to accomplish the parameter diminution for the trained neural networks. Specifically, we remove the feedback loop from the compression procedure pipeline and filter the neural connections in the different regions, to reduce the running complexity. The customized weight-level pruning method combines the pruning and fine-tuning process, as shown in Fig.~\ref{Weight-level slimming}. 

Pruning is applied to both the encoder and decoder respectively, because the corresponding parameters' distribution may be significantly different (encoder and decoder have different functions and number of parameters). The parameters with insignificant magnitude are advised to be pruned since the neurons firing together wire together. After the trained neural network is pruned with a certain ratio, the pruned neural network is retrained by the fine-tuning dataset to compensate for the degradation derived by pruning. We keep the pruned neural connections activated when calculating the gradients for back-propagation and then set the gradients of the pruned parameters to 0 when updating the parameters. Therefore, the remained parameters are updated by the fine-tuning process to smooth the pruned neural network while the pruned parameters are kept to be 0. It also provides insights into whether the pruned neural connections are not essential. If the pruning is applied appropriately, these gradients of the pruned parameters should not have a very critical magnitude when the training process is saturated. The pruned parameters with a significant gradient will be reactivated to avoid pruning the vital connections, which is proved biologically to damage the neural activity critically and the critical synapse is often enhanced \cite{faust2021mechanisms}. Otherwise, the gradient of pruned parameters will be kept to 0. Moreover, the compensation provided by the fine-tuning process is leaked by the gradient of pruned parameters and the degradation should be minimized if the pruned parameters is not essential. 

Therefore, neural networks should be firstly designed with large-scale parameter space and then slimmed by weight-level pruning, rather than proposing a compact design. Efficiency is a relative concept to the tasks the neural networks face, which weight-level pruning is exploited to balance with. The trade-off depends on whether the improved performance is essential compared with the computation cost derived. The paper \cite{hooker2019compressed} also discusses the effect of the pruning on the overall performance of neural networks. 
\section{Simulation results}
\label{Simulation results}
MSE is a key performance metric that evaluates the distance between the actual channel and the estimate of channel for each resource element in the slot, which is defined as  
\begin{equation}
    \mathrm{MSE}(\mathbf{\hat{H}}, \mathbf{H}) = \frac{1}{N_f N_s}\sum_{i=1}^{N_f} \sum_{j=1}^{N_s} {\left\Arrowvert\hat{H}_{ij} - H_{ij}\right\Arrowvert_{2}^{2}} , 
\end{equation}
Where $H_{ij}$ is the exact channel at subcarrier $i$ and OFDM symbol $j$ and $\hat{H}_{ij}$ is the corresponding estimate. The Denoising gain (DG) measured in dB is another performance metric which evaluates the ability of the neural network to improve the quality of the LS estimate. Equ.~(\ref{Denoise gain}) is a modified version of equ.~(2) in \cite{van2020optcomnet} to ensure a positive gain in dB when the channel estimate is improved by the neural network. 
\begin{equation}
    \mathrm{DG} = 10\log\frac{\Arrowvert \mathbf{\hat{H}_{LS}}-\mathbf{H}\Arrowvert_{2}^{2}}{\Arrowvert \mathbf{\hat{H}_{Channelformer}}-\mathbf{H}\Arrowvert_{2}^{2}} \ (\mathrm{dB}) . 
\label{Denoise gain}
\end{equation}

The bit error ratio (BER) is also a performance metric to evaluate the system performance. The offline training datasets for both the non-attention solutions and the attention-based solutions are generated by the hyperparameters from Table.~\ref{Parameters for training dataset generation}. The major difference between offline and online training relates to the regression label. For offline training, the channel matrix of the complete slot $\in \mathbb{R}^{{N_f} \times {N_s} \times 2}$ and is used for InterpolateNet, HA02, offline Channelformer and ReEsNet. The regression label for the TR method is the actual channel matrix at the pilot positions. The regression label for the online-trained Channelformer method is the channel matrix at the pilot symbols $\in \mathbb{R}^{N_{pilot}{N_f} \times 2}$. The hyperparameters for training ReEsNet and InterpolateNet are extracted from the original papers \cite{li2019deep} \cite{luan2021low}. To average out the Monte Carlo effects for all the simulation plots, each point is measured by 5000 independent channel realizations. 
\begin{table}[htbp]
\caption{Offline training hyperparameters and model size information}
\begin{center}
\begin{tabular}{|c|c|c|c|c|c|}
\hline
\textbf{}& \textbf{Online (Offline) Channelformer}& \textbf{HA02}& \textbf{InterpolateNet}& \textbf{ReEsNet}& \textbf{TR}\\
\hline
\textbf{Number of parameters}& 32,069 (117,659)& 105,607& 9,442& 53,000& 31,829\\
\hline
\textbf{Optimizer}& Adam (Adam)& Adam& Adam& Adam& Adam\\
\hline
\textbf{Maximum epoch}& 20 (100)& 100 & 100& 100& 100\\
\hline
\textbf{Initial learning rate (lr)}& 0.002 (0.002)& 0.002& 0.001& 0.001& 0.002\\
\hline
\textbf{Drop period for lr}& every 10 (every 50)& every 10& every 20& None& every 20\\
\hline
\textbf{Drop factor for lr}& 0.5 (0.5)& 0.5& 0.5& None& 0.5\\
\hline
\textbf{Minibatch size}& 128 (128)& 128& 128& 128& 128\\
\hline
\textbf{L2 regularization}& 1e-7 (1e-7)& 1e-7& 1e-7& 1e-7& 1e-7\\
\hline
\end{tabular}
\label{Offline training parameters}
\end{center}
\end{table}

The loss function deployed for InterpolateNet and ReEsNet is MSE, which is consistent with \cite{li2019deep} \cite{luan2021low}. By implementing the MSE loss function, the model converges rapidly with the drawback that the outlier samples in the training dataset can degrade the trained neural networks significantly. Compared with MSE, MAE is more robust to outlier samples with compromise on the convergence speed. For practical training, training scenarios with low SNR and time-varying channels, outlier samples exist in the training dataset. For online training, the neural network should converge rapidly to reduce the training latency. Therefore, the loss function employed for Channelformer, TR and HA02 is the Huber loss \cite{huber1965robust} defined in equ.~(\ref{huber}) with the transition $\delta$ set to 1. 
\begin{equation}
 L_{\delta}(a) =
\begin{cases}
\frac{1}{2}a^2& \text{if $|a| \leq \delta$}\\
\delta\left(|a| - \frac{1}{2}\delta\right)& \text{otherwise}
\end{cases} . 
\label{huber}
\end{equation}

It is a combination of MSE and MAE, which provides a compromise between the outlier impact and the convergence speed of the training process. The model is trained on a single NVIDIA GeForce RTX 2080 Super with Max-Q Design using MATLAB 2020a. For the neural networks updating the parameters online, only denoising and frequency interpolation can be achieved because time interpolation needs the noise-free channel matrix at the data OFDM symbols as mentioned. The average execution time of each method is provided in Table.~\ref{Execution time of each methods}. 
\begin{table}[htbp]
\caption{Parameters for offline-training dataset generation}
\begin{center}
\begin{tabular}{|c|c|c|c|c|}
\hline
\textbf{SNR values}& \textbf{Doppler frequency}& \textbf{Training dataset size}& \textbf{Channel Type}\\
\hline
From 5dB to 25dB& From 0Hz to 97Hz& 125,000 (95\% for training and 5\% for validation)& ETU\\
\hline
\end{tabular}
\label{Parameters for training dataset generation}
\end{center}
\end{table}
\begin{table}[htbp]
\caption{Average execution time of the deployed methods in milliseconds}
\begin{center}
\begin{tabular}{|c|c|c|c|c|c|c|c|}
\hline
\textbf{LS}& \textbf{DD-CE}& \textbf{1D (2D) FD-MMSE}& \textbf{Online (Offline) Channelformer}& \textbf{HA02}& \textbf{InterpolateNet}& \textbf{ReEsNet}& \textbf{TR}\\
\hline
0.5954& 15.8698 & 1.2260 (3.9597)& 20.5003 (31.0372)& 23.4283& 10.3835& 8.9523& 11.5869\\
\hline
\end{tabular}
\label{Execution time of each methods}
\end{center}
\end{table}
\subsection{MSE performance over the extended SNR range and Doppler shift}
\label{MSE performance over the extended SNR range}
We evaluate the MSE performance over an extended SNR range compared with the training dataset, to investigate the generalization capability for the unseen SNR values. The test dataset is generated on the ETU channel with extended SNR range from -10dB to 30dB and Doppler shift from 0Hz to 97Hz. Each SNR is tested with 5000 independent channel realizations. 
\begin{figure}[htbp]
\centering
\subfloat[MSE performance over the extended SNR range \label{MSE performance over the extended SNR}]{%
       \includegraphics[width=0.5\linewidth]{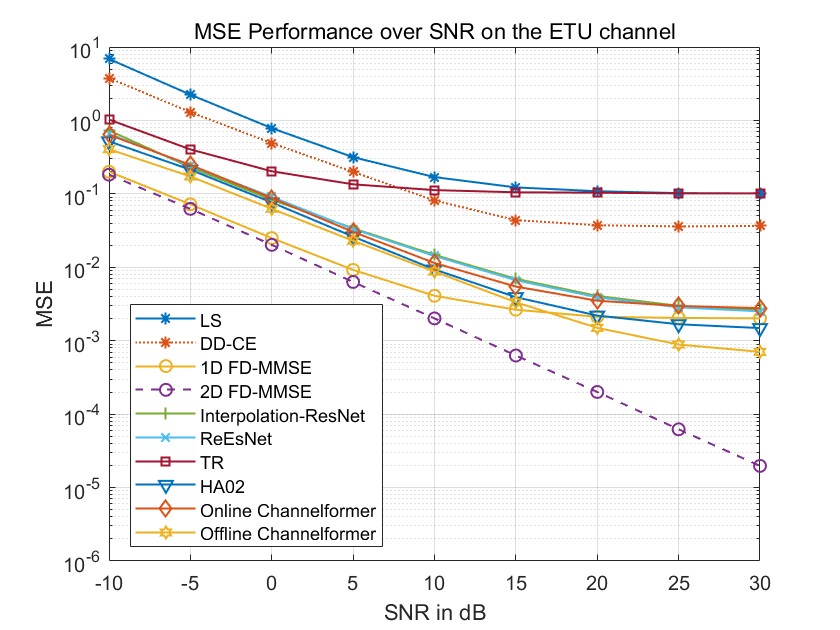}}
\hfill
\subfloat[MSE performance over the extended Doppler shift range \label{MSE performance over the extended Doppler shift}]{%
        \includegraphics[width=0.5\linewidth]{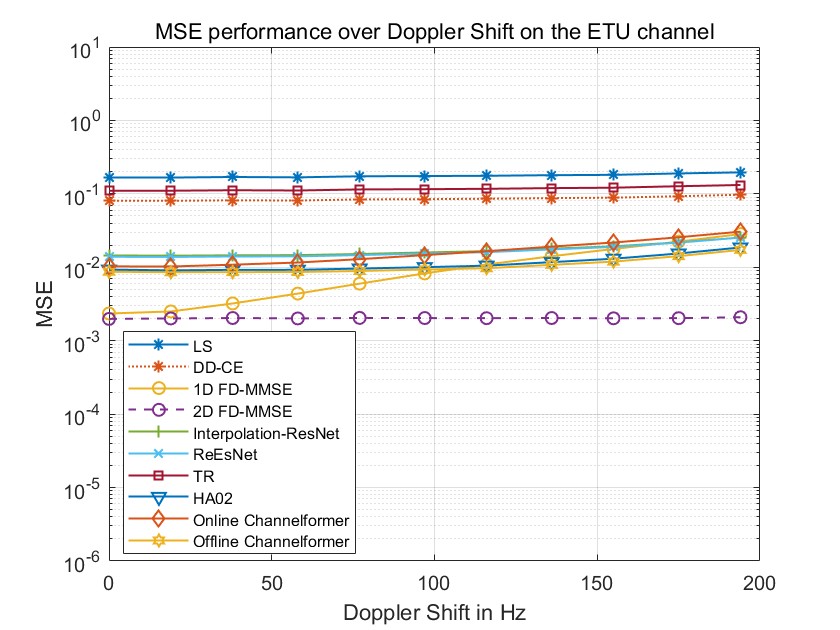}}
\caption{MSE performance of all channel estimation methods on the ETU channel}
\label{MSE performance}
\end{figure}

Fig.~\ref{MSE performance over the extended SNR} compares the MSE performance of each method over the extended SNR range. As in Fig.~\ref{MSE performance over the extended SNR}, the offline Channelformer outperforms all the other methods. Compared with HA02 and offline Channelformer, the MSE of online Channelformer increases by 0.0012 and 0.0020 for 30dB SNR respectively. However, it outperforms both InterpolateNet and ReEsNet for the whole SNR range. It should be noted that online Channelformer and TR are trained with part of channel information and interpolated by the linear method, which should degrade performance compared with offline neural networks trained with the complete channel information. However, the conventional neural networks are still worse than offline Channelformer and HA02. These results also show that the channel prediction of InterpolateNet and ReEsNet is not as precise when compared with the attention-based neural networks. Finally, we see that the performance of TR is significantly worse than other solutions and the 2D FD-MMSE method provides the most precise estimate among these methods. Compared with LS method, the DD-CE method has only a slightly improved performance. 

We also evaluate the MSE performance of each method over the extended Doppler shift range compared with the training dataset, to investigate the generalization capability to the Doppler shift. The test dataset is generated on the ETU channel with the condition that the SNR = 10dB and the Doppler shift is now from 0Hz to 194Hz (mobile speed from 0km/h to 100km/h). When the Doppler shift is set to 100Hz, it means that the maximum Doppler shift for each channel realization is randomly selected from 0Hz to 100Hz. Fig.~\ref{MSE performance over the extended Doppler shift} compares the MSE performance of each method over the extended Doppler shift range. As in Fig.~\ref{MSE performance over the extended Doppler shift}, the HA02 and offline Channelformer methods outperform the InterpolateNet and ReEsNet for the whole Doppler shift range. For maximum Doppler shifts below 125Hz, the online Channelformer method outperforms the conventional neural networks but the performance reduces as the maximum Doppler shift increases. At the high Doppler shift range, the online Channelformer achieves similar MSE results to the 1D FD-MMSE method but degrades compared with the InterpolateNet and ReEsNet. The effect of Doppler shift impacts the channel variation in the time domain and the linear method exploits absolutely no information of the Doppler shift to achieve the time domain interpolation. Therefore, HA02 and offline Channelformer can outperform 1D FD-MMSE at the high Doppler shift range when the effects of Doppler shift are significant, while non-attention solutions (InterpolationNet and ReEsNet) are worse than the 1D FD-MMSE method. Moreover, TR achieves much poorer performance when compared with the other solutions while the 2D FD-MMSE method still has the best MSE performance among these methods. Compared with LS method, the DD-CE method also provides an slightly improved performance. 
\subsection{Complexity reduction: customized weight-level pruning}
\label{Complexity reduction: weight-level slimming and denoise gain}
\begin{figure}[htbp]
\centering
\subfloat[Customized weight-level pruning \label{MSE performance and weight-level slimming}]{%
        \includegraphics[width=0.5\linewidth]{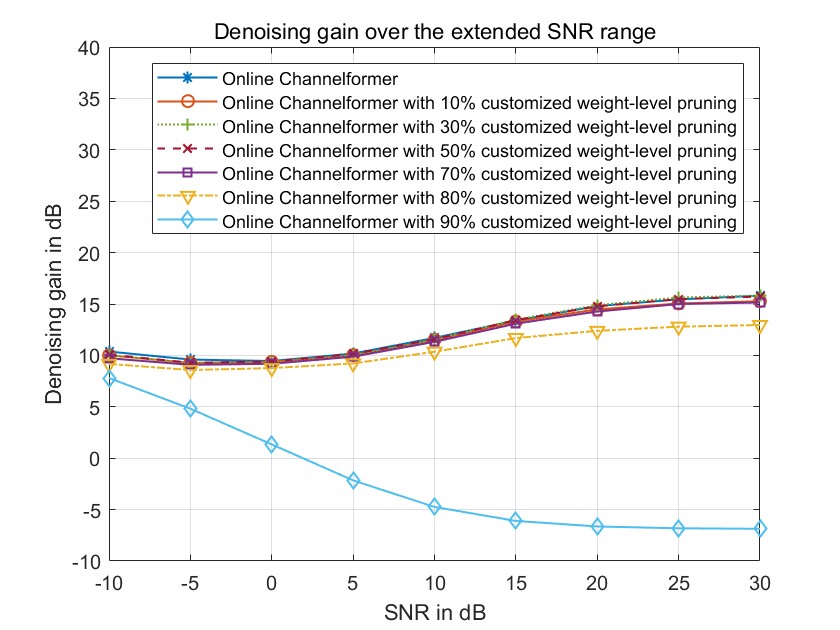}}
\hfill
\subfloat[Weight-level pruning without fine-tuning \label{Weight-level pruning without fine-tuning process}]{%
       \includegraphics[width=0.5\linewidth]{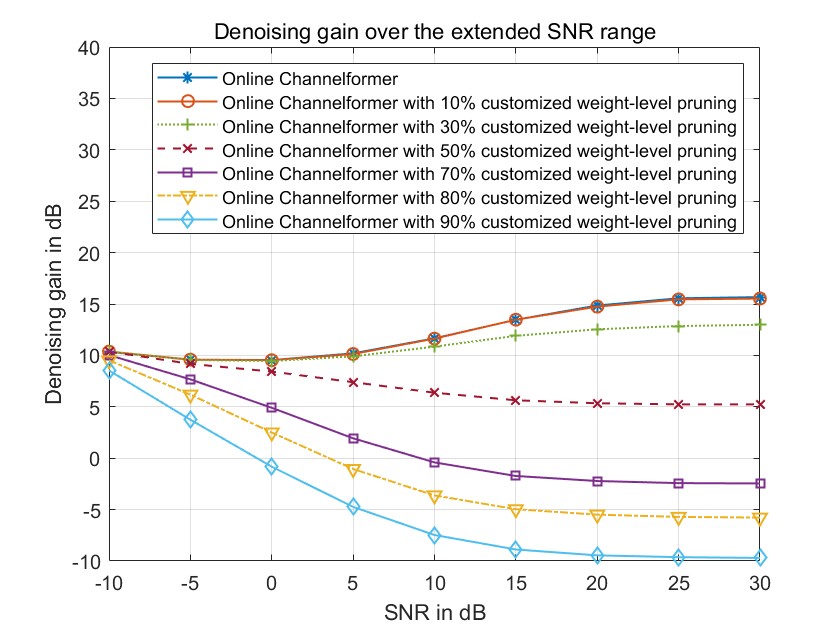}}
\caption{Denoising gain performance when tested on the ETU channel}
\label{weight-level slimming and denoise gain}
\end{figure}

To investigate the impact of the customized weight-level pruning on the generalization to the extended SNR range, we apply that to the online Channelformer to reduce the complexity further. For the fine-tuning process, most training settings and hyperparameters from Table.~\ref{Offline training parameters} are retained. The maximum epoch is changed to 10 with lr = 0.001, the batch size is reduced to 32 and the training dataset is composed of 15,000 samples. The test dataset for Fig.~\ref{weight-level slimming and denoise gain} is generated identically with Section.~\ref{MSE performance over the extended SNR range}. Each SNR is also tested with 5000 channel realizations to average out Monte Carlo effects. 

Fig.~\ref{MSE performance and weight-level slimming} compares the DG of each pruning ratio for customized weight-level pruning over the extended SNR range. The performance of each customized-slimmed Channelformer is almost unchanged while the pruning ratio $\leq$ 70\%. However, 80\% pruning leads to a 7dB gain lost on average while 90\% pruning degrades the performance completely. Fig.~\ref{Weight-level pruning without fine-tuning process} provides the results that apply the pruning to the trained neural network without fine-tuning. Compared with customized weight-level pruning, the performance degrades significantly at the pruning ratio of 30\% because the pruned neural network is not retrained to compensate for the loss. The 70\% customized weight-level pruned online Channelformer has a 0.3dB degradation approximately compared with the online Channelformer without pruning and the number of parameters is 9,620. 
\begin{figure}[htbp]
\centering
\subfloat[Customized weight-level pruned Channelformer \label{MSE performance of slimmed Channelformer (Doppler shift)}]{%
       \includegraphics[width=0.5\linewidth]{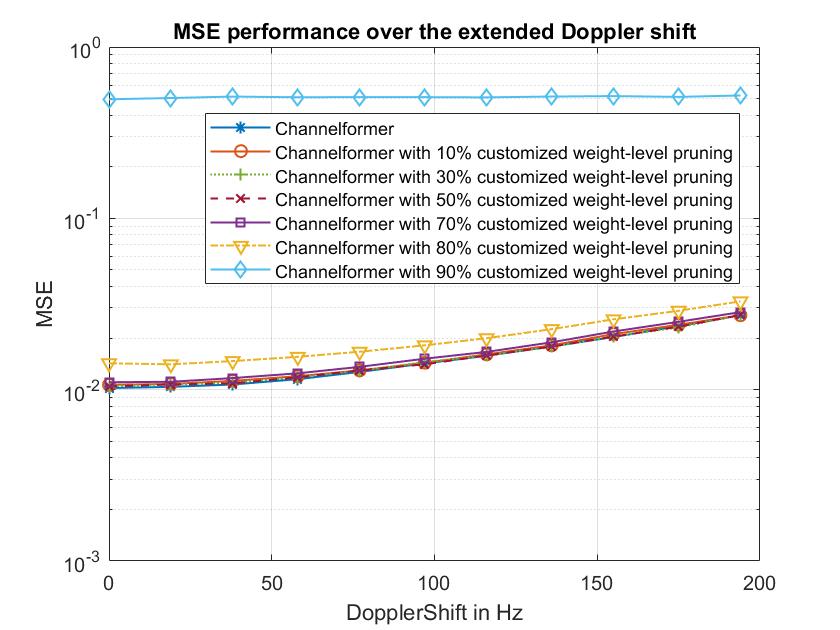}}
\hfill
\subfloat[Weight-level pruned Channelformer without fine-tuning \label{MSE performance of pruned Channelformer (Doppler shift)}]{%
        \includegraphics[width=0.5\linewidth]{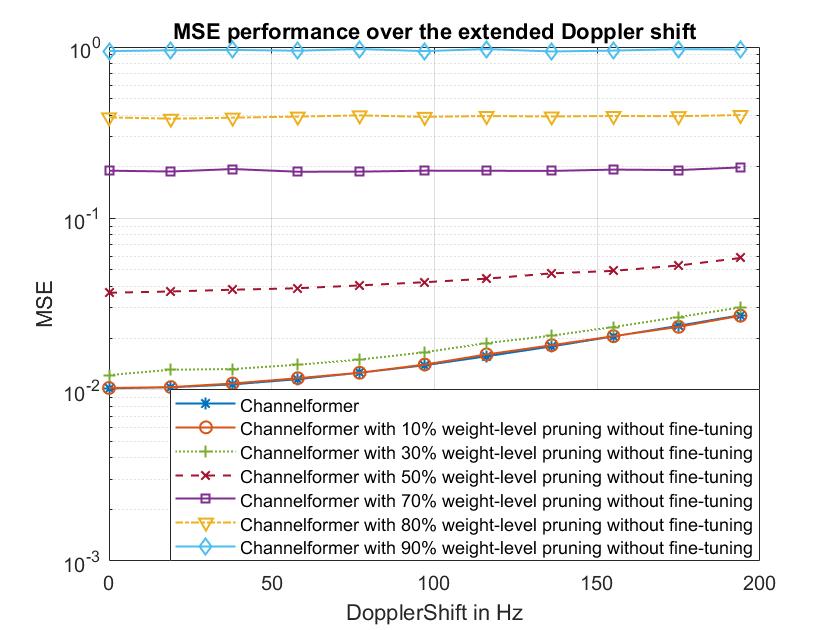}}
\caption{Generalization to the extended Doppler shift range}
\label{Generalization to the extended Doppler shift range}
\end{figure}

\begin{figure}[htbp]
\centering
\subfloat[Denoising gain performance on the extended SNR \label{Denoising gain performance on the extended SNR range}]{%
       \includegraphics[width=0.5\linewidth]{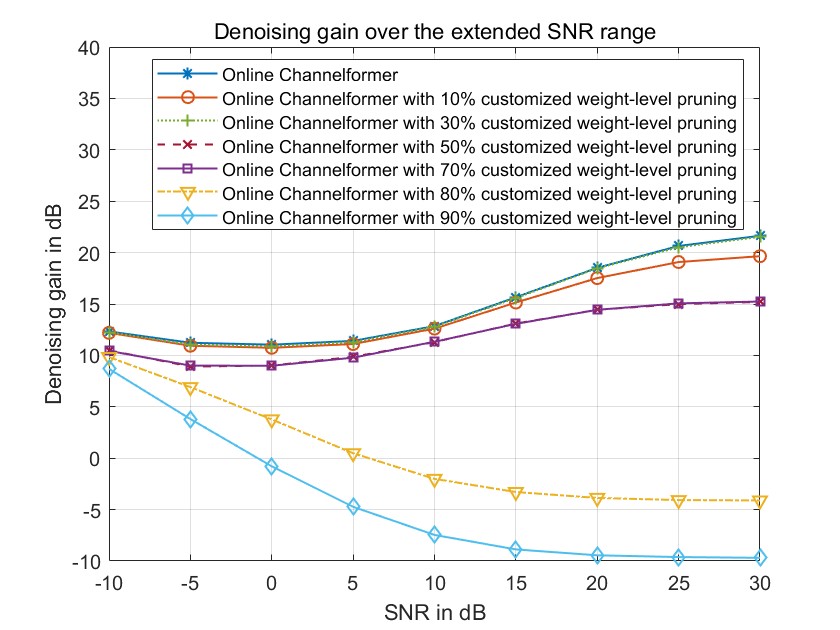}}
\hfill
\subfloat[MSE performance on the extended Doppler shift \label{Denoising gain performance on the extended Doppler shift range}]{%
        \includegraphics[width=0.5\linewidth]{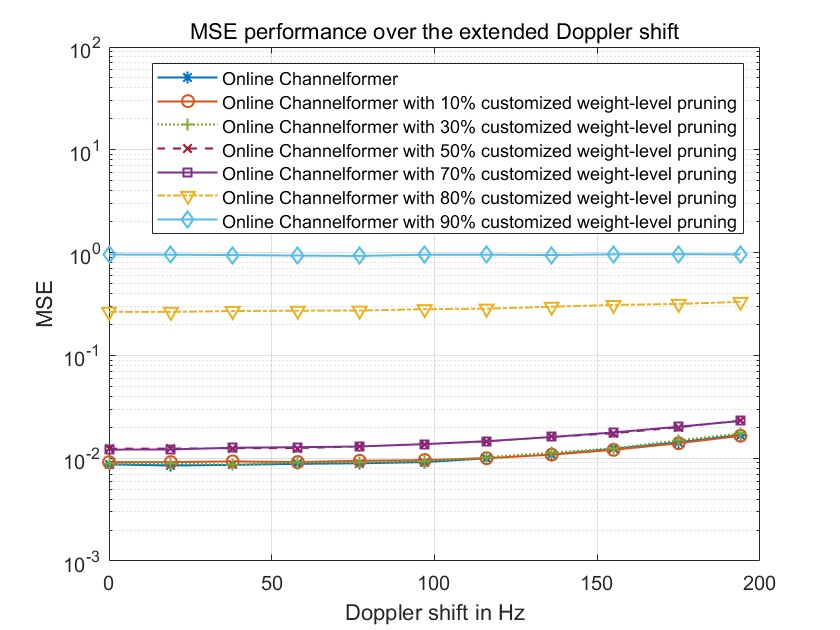}}
\caption{Customized weight-level pruned offline Channelformer when tested on the ETU channel}
\label{Customized weight-level pruned offline Channelformer}
\end{figure}

To investigate the impact of the customized weight-level pruning on the generalization to the extended Doppler shift range, Fig.~\ref{Generalization to the extended Doppler shift range} compare the MSE performance of slimmed online Channelformer. The test dataset is identical to the test dataset in Section.~\ref{MSE performance over the extended Doppler shift}. Even 70\% pruning only gives an extremely slight increase in MSE performance, which is almost same as the online Channelformer without pruning. Therefore, 70\% customized weight-level pruned Channelformer is observed to have a robust performance to the extended SNR and Doppler shift, which should be sufficient for the practical implementation. It also indicates that redundant parameters do not significantly affect the neural network's robust performance with the extended SNR and Doppler shift ranges. Therefore, the redundant architecture may have resistance to the unknown changes because the customized weight-level pruning removes the redundant parameters and the performance is retained for a pruning ratio under 70\% when compared with the complete online Channelformer. This is in contrast to findings in \cite{zhang2017defense} for the field of image processing, as the customized weight-level pruning removes the redundant parameters. We also simulate the offline Channelformer applied by the customized weight-level pruning shown in Figure.~\ref{Customized weight-level pruned offline Channelformer}. 
\subsection{BER performance over the extended SNR range}
\begin{figure}[htbp]
\centering
\subfloat[BER performance on the extended SNR (ETU channel) \label{BER performance over the extended SNR}]{%
       \includegraphics[width=0.5\linewidth]{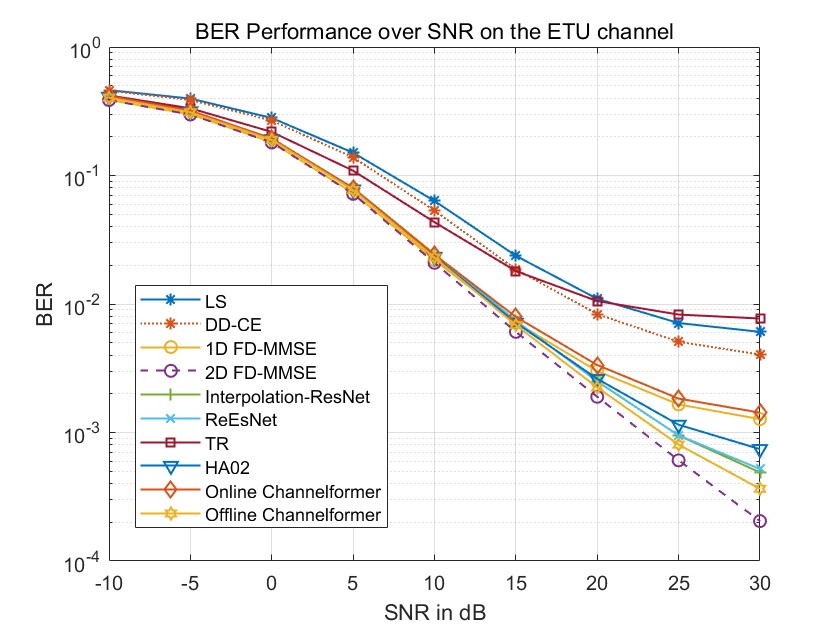}}
\hfill
\subfloat[BER performance on the extended SNR (customized channel) \label{BER performance on the varied channel with the extended SNR}]{%
        \includegraphics[width=0.5\linewidth]{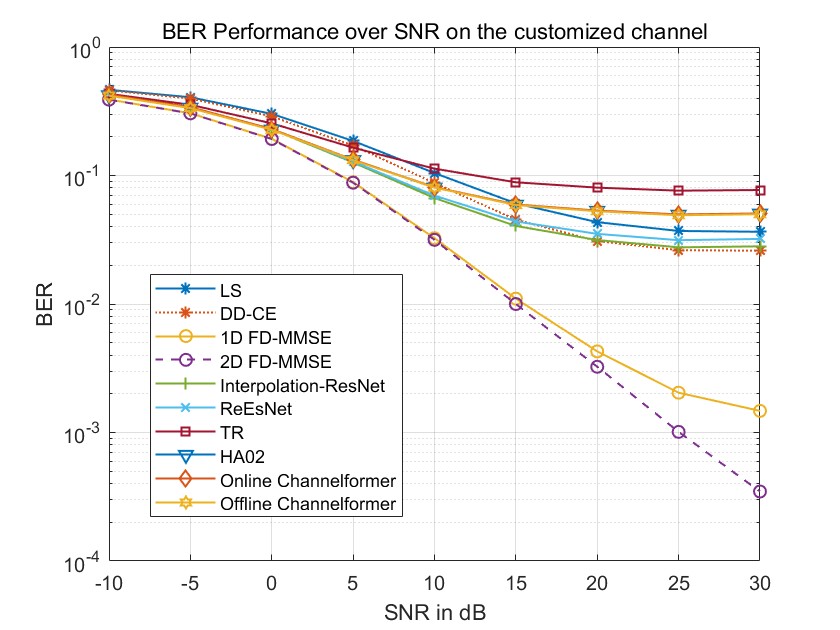}}
\caption{BER performance on the ETU and customized channels}
\end{figure}

We also evaluate the BER performance of each method over the extended SNR range. The test dataset for Fig.~\ref{BER performance over the extended SNR} is generated identically with Section.~\ref{MSE performance over the extended SNR range}. 

Fig.~\ref{BER performance over the extended SNR} compares the BER performance of each method and the offline Channelformer method achieves the best performance. It also indicates a possible opposed result compared with Fig.~\ref{MSE performance over the extended SNR} that both InterpolateNet and ReEsNet can outperform 1D FD-MMSE and online Channelformer for the SNR above 20dB. The reason is that, for online Channelformer and 1D FD-MMSE, the prediction at the pilot symbols is extremely precise althrough the capability of time interpolation for the linear method is much worse than that achieved by neural networks (InterpolateNet and ReEsNet). Therefore, the MSE performance in Fig.~\ref{MSE performance over the extended SNR} is still better than InterpolateNet and ReEsNet. However, the linear interpolation introduces the degradation as mentioned, compared with the neural network trained using the complete set of channel coefficients for the whole slot. The BER performance depends on the precision of the channel matrix prediction at all of the data symbols. InterpolateNet and ReEsNet are trained with the channel matrix of the whole slot while online Channelformer and 1D FD-MMSE only exploit the channel matrix at the pilot symbols. When predicting the channel for the data symbols, time interpolation achieved by linear interpolation is significantly worse than that achieved by neural networks. Therefore, the BER performance of InterpolateNet and ReEsNet is superior to 1D FD-MMSE and online Channelformer for the SNR above 15dB. Compared with the InterpolateNet and ReEsNet, the offline Channelformer trained with the same labels outperforms the InterpolateNet and ReEsNet. The gain achieved by neural network methods would be limited under the fair comparison that the neural networks are also trained by the channel matrix at the pilot symbols and all approaches use the same time interpolation method. It should be noted that online training cannot access the channel matrix at the data symbols. Moreover, the BER of the 2D FD-MMSE method is the lowest among these methods. When the trained neural networks are tested on the channel defined in Table.~\ref{Online channel}, the degradation of each neural network is clearly observed in Fig.~\ref{BER performance on the varied channel with the extended SNR}, which motivates the application of online training. 
\begin{table}[htbp]
\caption{The PDP of the customized Channel}
\begin{center}
\begin{tabular}{|c|c|}
\hline
\textbf{Path Delays}& [0, 30, 200, 300, 500, 1500, 2500, 5000, 7000, 9000] ns\\
\hline
\textbf{Average Path Gain}& [-1.0, 0, 0, -1.0, -2.0, -1.0, -1.0, -1.5, -3.0, -5.0] dB\\
\hline
\end{tabular}
\label{Online channel}
\end{center}
\end{table}
\subsection{Dynamic online adaptation}
We simulate the dynamic adaptation of online Channelformer, which is synchronous with the transmission of the data. Different from collecting sufficient samples for training, the neural network tracks the channel by training with a batch of 3 online samples before predicting the channel matrix (each iteration of online training takes 99.8ms for the complete online Channelformer and approximately 29.9ms for the 70\% pruned online Channelformer). These times are too slow for real time implementation, but could be speeded up using dedicated hardware. The online Channelformers are trained offline first because starting with the initialized parameters just takes more time, and the pruned online Channelformer is fine-tuned by the online samples rather than the offline dataset. In the test period, the dynamic channel alternates to other candidate channels every 10000 channel realisations. The channel assemble involves ETU, the customized channel, EVA and an advanced channel with the extended SNR from 15dB to 25dB and the mobile speed is from 0km/h to 50km/h. The advanced channel is a SISO version of the channel deployed in \cite{narengerile2021deep} following 3GPP TS38.901 section 7.5 to generate the fast fading channel, representing a more realistic scenario. Each MSE sample is averaged over blocks of 50 channel realisations. The fine-tuning procedure of the customized weight-level pruning is substituted by the online training process. 
\begin{figure}[htbp]
\centering
\subfloat[Dynamic adaptation simulation \label{Dynamic adaptation}]{%
       \includegraphics[width=0.5\linewidth]{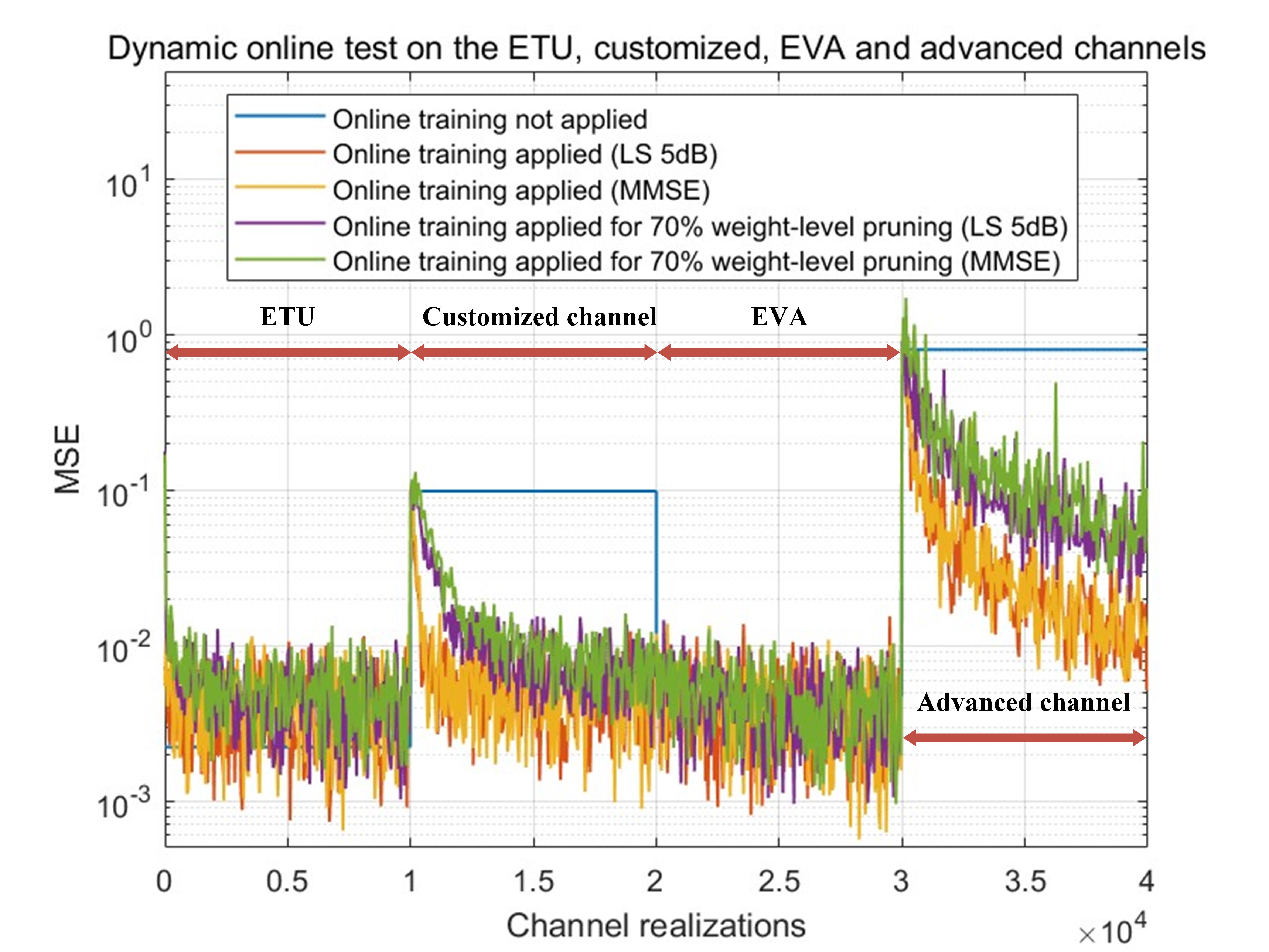}}
\hfill
\subfloat[Averaged label precision of both label pilot symbols \label{Estimate bias of the customized channel}]{%
        \includegraphics[width=0.5\linewidth]{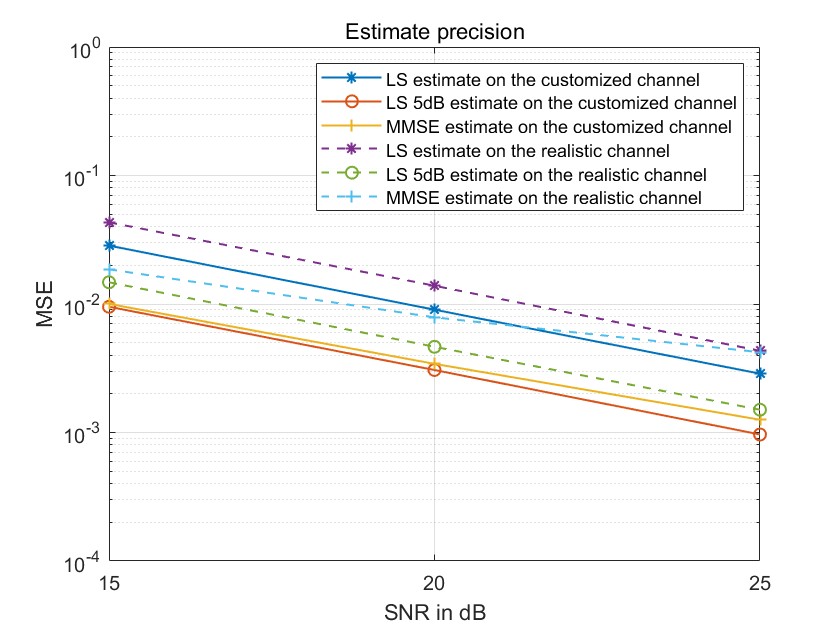}}
\caption{Online neural network simulation. The simulation channel alternatives among the ETU channel, the customized channel, EVA channel and the advanced channel}
\end{figure}

Fig.~\ref{Dynamic adaptation} gives the dynamic performance and the MSE for offline training only is averaged over each block of 10000 samples, as the neural network is not updated with the new channel information. When the channel is changed to the customized channel in Table.~\ref{Online channel}, the MSE converges to a stable value of 0.01 within first 50 samples while the online training process is synchronous with the transmission of the data. The 70\% customised weight-level pruned online Channelformer converges to a stable MSE loss of 0.02 within 550 samples, which is slower than the complete online Channelformer because 70\% neural connections are pruned (reduced capacity). When the channel is alternated to the EVA channel, the MSE is equal for both the untrained online Channelformers and the trained online Channelformers. It proves that online Channelformer can generalize to EVA channel and the customized weight-level pruning has negligible impact on that generalization. When the channel is alternated to the fast fading channel \cite{narengerile2021deep}, the considerable mismatch in channel modelling leads to a significant increase in MSE at the start. The stable MSE of both unpruned neural networks decreases to 0.02 within 5000 channel realizations while the pruned neural networks converge to approximately 0.035 because pruning reduces their capability to adapt to previously unseen channels. 
\subsection{Attention analysis}
\label{Attention}
\begin{figure}[htbp]
\centering
    \subfloat[Head 1, channel 1]{\includegraphics[width=0.5\linewidth]{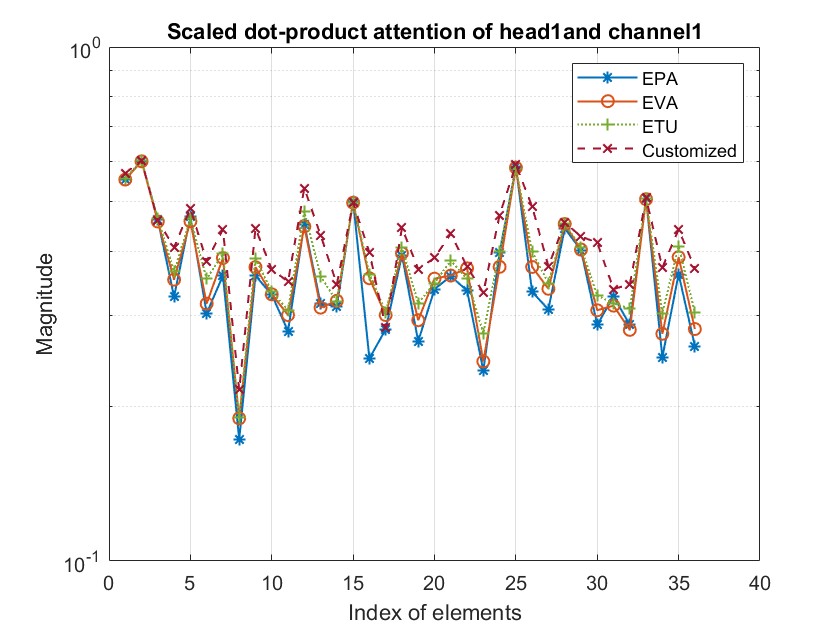}}\hfill 
    \subfloat[Head 1, channel 2]{\includegraphics[width=0.5\linewidth]{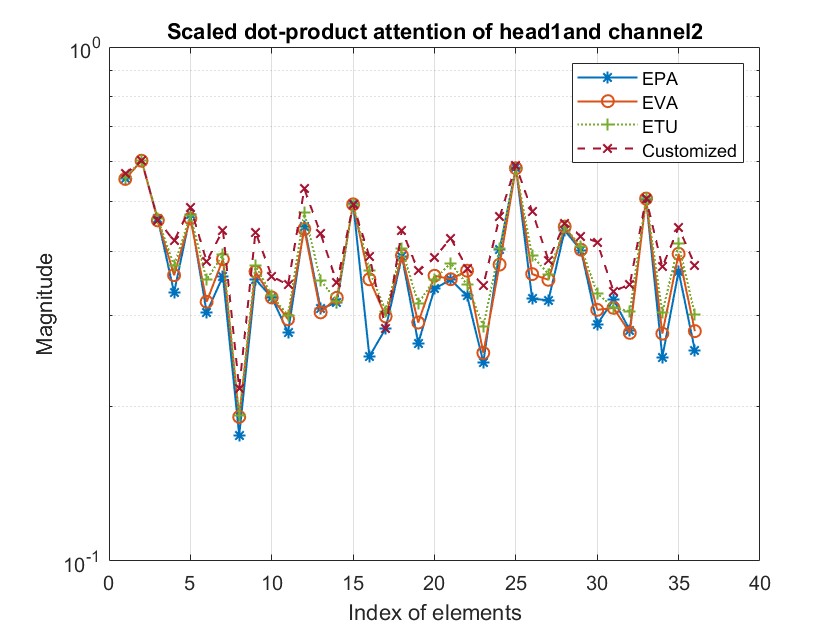}}\hfill 
    \subfloat[Head 2, channel 1]{\includegraphics[width=0.5\linewidth]{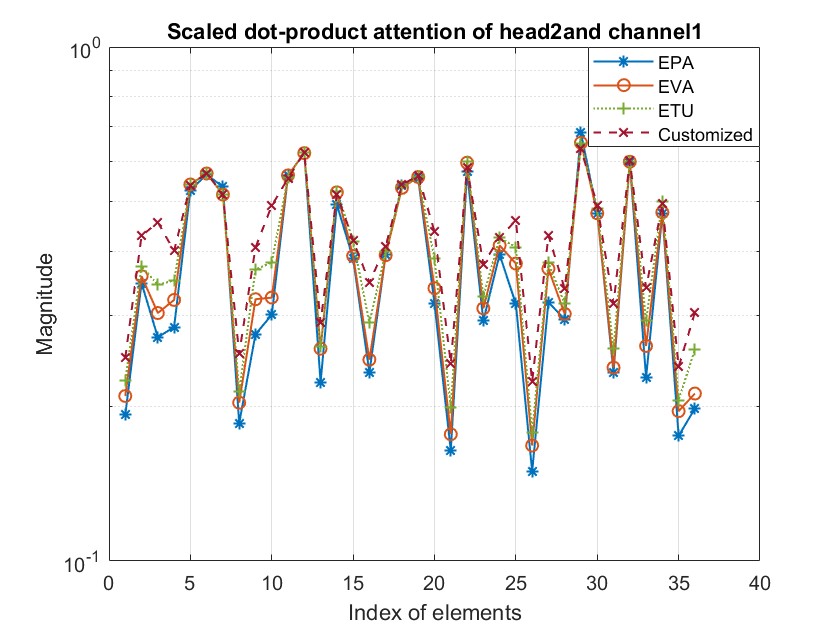}}\hfill 
    \subfloat[Head 2, channel 2]{\includegraphics[width=0.5\linewidth]{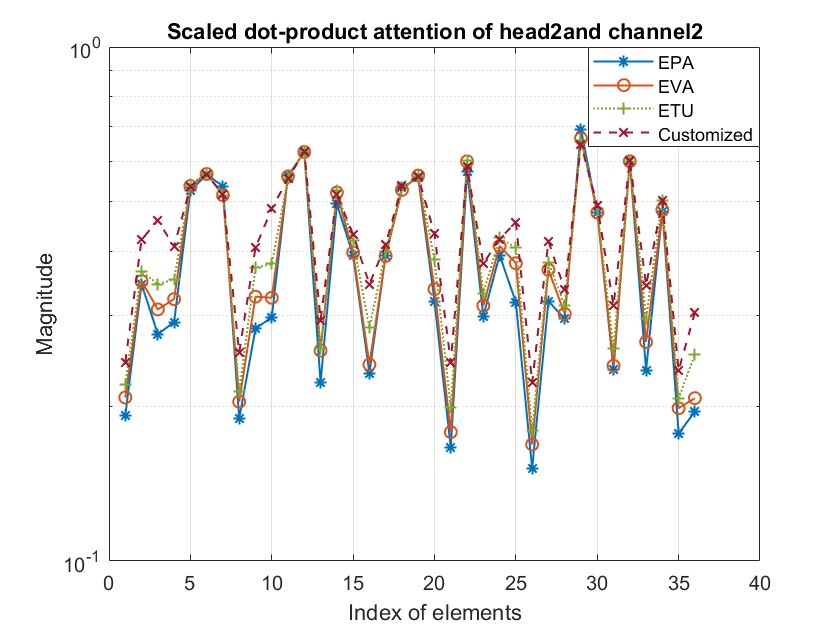}}\hfill 
    \caption{Mean of scaled dot-product attention for EPA, EVA, ETU and the customized channels}
    \label{Attention analyse}
\end{figure}

The attention mechanism is employed to achieve the input precoding forming the encoded features. We collect the scaled dot-product attention defined by equ.~(\ref{scaled dot-product attention}) for each head when the online Channelformer is tested on the EPA, EVA, ETU and customized channels.  Fig.~\ref{Attention analyse} compares the mean of the corresponding attention's magnitude. The attention for each head $\in \mathbb{R}^{\left(\frac{N_f}{L_s}\right) \times 2}$ has 2 channels, which is the 2\textsuperscript{nd} dimension of the attention. 

Fig.~\ref{Attention analyse} provides the mean of the scaled dot-product attention when predicting the channel matrix for the EPA, EVA, ETU and customized channels. It is not uniform, which proves that the input precoding is achieved as expected. The scaled dot-product attention for the EPA, the EVA channel and the ETU channel are almost similar. For head 1, the 1\textsuperscript{th}, 2\textsuperscript{th} and 25\textsuperscript{th} elements are more important than other elements while the impact of the 8\textsuperscript{th} element is negligible. For head 2, the 29\textsuperscript{th} element has the most significant magnitude. For the customized channel, the 1\textsuperscript{th}, 2\textsuperscript{th} and 25\textsuperscript{th} elements are more important than other elements for head 1 and the 12\textsuperscript{th}, 22\textsuperscript{th} and 29\textsuperscript{th} element has the most significant magnitude for head 2. It also proves that the scaled dot-product attention's magnitude differs for different heads because the multi-head attention focuses on the different parts of the signal in different heads. Moreover, the mean of the scaled dot-product attention is almost same for channel 1 and channel 2. 
\section{Conclusion}
\label{Conclusion}
We have proposed an encoder-decoder neural architecture called Channelformer for wireless channel estimation in outdoors downlink scenarios. Channelformer integrates the self-attention mechanism to achieve input precoding in the encoder. The encoded features are processed by the residual convolutional architecture to predict the channel matrix in the decoder. We study the performance of both offline-trained and online-trained variants of the Channelformer network, and investigate the complexity reduction by customized weight-level pruning and apply that to the online training algorithm proposed. From the simulation results, Channelformer outperforms the other baseline neural solutions on the extended SNR range and Doppler shift when tested on the ETU channel. For complexity reduction, the 70\% weight-level pruned online Channelformer (9,620 parameters) retains an almost identical performance compared with the complete neural network when tested on extended channel conditions. For online training, the proposed algorithm is effective when tested in a practical environment. The input precoding achieved by the self-attention mechanism is also analysed. 
%
\section*{Acknowledgment}

The authors gratefully acknowledge the funding of this research by Huawei. For the purpose of open access, the author has applied a Creative Commons Attribution (CC BY) licence to any Author Accepted Manuscript version arising from this submission. To support reproducibility, the simulation code can be downloaded at \url{https://doi.org/10.7488/ds/3801} and \url{https://github.com/dianixn/Channelformer}. 

\ifCLASSOPTIONcaptionsoff
  \newpage
\fi

\bibliographystyle{IEEEtran}

\bibliography{Reference}

\begin{thebibliography}{10}
\providecommand{\url}[1]{#1}
\csname url@samestyle\endcsname
\providecommand{\newblock}{\relax}
\providecommand{\bibinfo}[2]{#2}
\providecommand{\BIBentrySTDinterwordspacing}{\spaceskip=0pt\relax}
\providecommand{\BIBentryALTinterwordstretchfactor}{4}
\providecommand{\BIBentryALTinterwordspacing}{\spaceskip=\fontdimen2\font plus
\BIBentryALTinterwordstretchfactor\fontdimen3\font minus
  \fontdimen4\font\relax}
\providecommand{\BIBforeignlanguage}[2]{{%
\expandafter\ifx\csname l@#1\endcsname\relax
\typeout{** WARNING: IEEEtran.bst: No hyphenation pattern has been}%
\typeout{** loaded for the language `#1'. Using the pattern for}%
\typeout{** the default language instead.}%
\else
\language=\csname l@#1\endcsname
\fi
#2}}
\providecommand{\BIBdecl}{\relax}
\BIBdecl

\bibitem{dang2020should}
S.~Dang, O.~Amin, B.~Shihada, and M.-S. Alouini, ``What should 6{G} be?''
  \emph{Nature Electronics}, vol.~3, no.~1, pp. 20--29, 2020.

\bibitem{shafi20175g}
M.~Shafi, A.~F. Molisch, P.~J. Smith, T.~Haustein, P.~Zhu, P.~De~Silva,
  F.~Tufvesson, A.~Benjebbour, and G.~Wunder, ``5{G}: A tutorial overview of
  standards, trials, challenges, deployment, and practice,'' \emph{IEEE Journal
  on Selected Areas in Communications}, vol.~35, no.~6, pp. 1201--1221, 2017.

\bibitem{edfors1998ofdm}
O.~Edfors, M.~Sandell, J.-J. Van~de Beek, S.~K. Wilson, and P.~O. Borjesson,
  ``{OFDM} channel estimation by singular value decomposition,'' \emph{IEEE
  Transactions on communications}, vol.~46, no.~7, pp. 931--939, 1998.

\bibitem{van1995channel}
J.-J. Van De~Beek, O.~Edfors, M.~Sandell, S.~K. Wilson, and P.~O. Borjesson,
  ``On channel estimation in {OFDM} systems,'' in \emph{1995 IEEE 45th
  Vehicular Technology Conference. Countdown to the Wireless Twenty-First
  Century}, vol.~2.\hskip 1em plus 0.5em minus 0.4em\relax IEEE, 1995, pp.
  815--819.

\bibitem{lottici2002channel}
V.~Lottici, A.~D'Andrea, and U.~Mengali, ``Channel estimation for
  ultra-wideband communications,'' \emph{IEEE Journal on Selected Areas in
  Communications}, vol.~20, no.~9, pp. 1638--1645, 2002.

\bibitem{letaief2019roadmap}
K.~B. Letaief, W.~Chen, Y.~Shi, J.~Zhang, and Y.-J.~A. Zhang, ``The roadmap to
  {6G: AI} empowered wireless networks,'' \emph{IEEE Communications Magazine},
  vol.~57, no.~8, pp. 84--90, 2019.

\bibitem{he2018deep}
H.~He, C.-K. Wen, S.~Jin, and G.~Y. Li, ``Deep learning-based channel
  estimation for beamspace mmwave massive {MIMO} systems,'' \emph{IEEE Wireless
  Communications Letters}, vol.~7, no.~5, pp. 852--855, 2018.

\bibitem{soltani2019deep}
M.~Soltani, V.~Pourahmadi, A.~Mirzaei, and H.~Sheikhzadeh, ``Deep
  learning-based channel estimation,'' \emph{IEEE Communications Letters},
  vol.~23, no.~4, pp. 652--655, 2019.

\bibitem{li2019deep}
L.~Li, H.~Chen, H.-H. Chang, and L.~Liu, ``Deep residual learning meets ofdm
  channel estimation,'' \emph{IEEE Wireless Communications Letters}, vol.~9,
  no.~5, pp. 615--618, 2019.

\bibitem{luan2021low}
D.~Luan and J.~Thompson, ``Low complexity channel estimation with neural
  network solutions,'' in \emph{WSA 2021; 25th International ITG Workshop on
  Smart Antennas}.\hskip 1em plus 0.5em minus 0.4em\relax VDE, 2021, pp. 1--6.

\bibitem{yan2019stat}
C.~Yan, Y.~Tu, X.~Wang, Y.~Zhang, X.~Hao, Y.~Zhang, and Q.~Dai, ``Stat:
  Spatial-temporal attention mechanism for video captioning,'' \emph{IEEE
  transactions on multimedia}, vol.~22, no.~1, pp. 229--241, 2019.

\bibitem{vaswani2017attention}
A.~Vaswani, N.~Shazeer, N.~Parmar, J.~Uszkoreit, L.~Jones, A.~N. Gomez,
  {\L}.~Kaiser, and I.~Polosukhin, ``Attention is all you need,'' in
  \emph{Advances in neural information processing systems}, 2017, pp.
  5998--6008.

\bibitem{devlin2018bert}
J.~Devlin, M.-W. Chang, K.~Lee, and K.~Toutanova, ``Bert: Pre-training of deep
  bidirectional transformers for language understanding,'' \emph{arXiv preprint
  arXiv:1810.04805}, 2018.

\bibitem{dosovitskiy2020image}
A.~Dosovitskiy, L.~Beyer, A.~Kolesnikov, D.~Weissenborn, X.~Zhai,
  T.~Unterthiner, M.~Dehghani, M.~Minderer, G.~Heigold, S.~Gelly \emph{et~al.},
  ``An image is worth 16x16 words: Transformers for image recognition at
  scale,'' \emph{arXiv preprint arXiv:2010.11929}, 2020.

\bibitem{pan2021channel}
J.~Pan, H.~Shan, R.~Li, Y.~Wu, W.~Wu, and T.~Q. Quek, ``Channel estimation
  based on deep learning in vehicle-to-everything environments,'' \emph{IEEE
  Communications Letters}, vol.~25, no.~6, pp. 1891--1895, 2021.

\bibitem{lu2021channel}
Q.~Lu, T.~Lin, and Y.~Zhu, ``Channel estimation and hybrid precoding for
  millimeter wave communications: a deep learning-based approach,'' \emph{IEEE
  Access}, vol.~9, pp. 120\,924--120\,939, 2021.

\bibitem{jiang2021dual}
P.~Jiang, C.-K. Wen, S.~Jin, and G.~Y. Li, ``Dual cnn-based channel estimation
  for {MIMO-OFDM} systems,'' \emph{IEEE Transactions on Communications},
  vol.~69, no.~9, pp. 5859--5872, 2021.

\bibitem{chen2020channel}
Z.~Chen, F.~Gu, and R.~Jiang, ``Channel estimation method based on transformer
  in high dynamic environment,'' in \emph{2020 International Conference on
  Wireless Communications and Signal Processing (WCSP)}.\hskip 1em plus 0.5em
  minus 0.4em\relax IEEE, 2020, pp. 817--822.

\bibitem{mashhadi2021pruning}
M.~B. Mashhadi and D.~G{\"u}nd{\"u}z, ``Pruning the pilots: Deep learning-based
  pilot design and channel estimation for {MIMO-OFDM} systems,'' \emph{IEEE
  Transactions on Wireless Communications}, 2021.

\bibitem{wang2018non}
X.~Wang, R.~Girshick, A.~Gupta, and K.~He, ``Non-local neural networks,'' in
  \emph{Proceedings of the IEEE conference on computer vision and pattern
  recognition}, 2018, pp. 7794--7803.

\bibitem{tekbiyik2021channel}
K.~Tekb{\i}y{\i}k, G.~K. Kurt, C.~Huang, A.~R. Ekti, and H.~Yanikomeroglu,
  ``Channel estimation for full-duplex {RIS}-assisted haps backhauling with
  graph attention networks,'' in \emph{ICC 2021-IEEE International Conference
  on Communications}.\hskip 1em plus 0.5em minus 0.4em\relax IEEE, 2021, pp.
  1--6.

\bibitem{luan2022attention}
D.~Luan and J.~Thompson, ``Attention based neural networks for wireless channel
  estimation,'' \emph{arXiv preprint arXiv:2204.13465}, 2022, to appear in IEEE
  VTC Spring 2022.

\bibitem{yang2021deep}
A.~Yang, P.~Sun, T.~Rakesh, B.~Sun, and F.~Qin, ``Deep learning based ofdm
  channel estimation using frequency-time division and attention mechanism,''
  \emph{arXiv preprint arXiv:2107.07161}, 2021.

\bibitem{gao2021attention}
J.~Gao, M.~Hu, C.~Zhong, G.~Y. Li, and Z.~Zhang, ``An attention-aided deep
  learning framework for massive {MIMO} channel estimation,'' \emph{IEEE
  Transactions on Wireless Communications}, 2021.

\bibitem{liu2012cost}
L.~Liu, C.~Oestges, J.~Poutanen, K.~Haneda, P.~Vainikainen, F.~Quitin,
  F.~Tufvesson, and P.~De~Doncker, ``The cost 2100 {MIMO} channel model,''
  \emph{IEEE Wireless Communications}, vol.~19, no.~6, pp. 92--99, 2012.

\bibitem{zheng2021online}
X.~Zheng and V.~K. Lau, ``Online deep neural networks for mmwave massive {MIMO}
  channel estimation with arbitrary array geometry,'' \emph{IEEE Transactions
  on Signal Processing}, vol.~69, pp. 2010--2025, 2021.

\bibitem{jha2021online}
N.~K. Jha and V.~K. Lau, ``Online downlink multi-user channel estimation for
  mmwave systems using bayesian neural network,'' \emph{IEEE Journal on
  Selected Areas in Communications}, vol.~39, no.~8, pp. 2374--2387, 2021.

\bibitem{yang2019deep}
Y.~Yang, F.~Gao, X.~Ma, and S.~Zhang, ``Deep learning-based channel estimation
  for doubly selective fading channels,'' \emph{IEEE Access}, vol.~7, pp.
  36\,579--36\,589, 2019.

\bibitem{mei2021low}
K.~Mei, J.~Liu, X.~Zhang, K.~Cao, N.~Rajatheva, and J.~Wei, ``A low complexity
  learning-based channel estimation for ofdm systems with online training,''
  \emph{IEEE Transactions on Communications}, vol.~69, no.~10, pp. 6722--6733,
  2021.

\bibitem{huang2018deep}
H.~Huang, J.~Yang, H.~Huang, Y.~Song, and G.~Gui, ``Deep learning for
  super-resolution channel estimation and doa estimation based massive {MIMO}
  system,'' \emph{IEEE Transactions on Vehicular Technology}, vol.~67, no.~9,
  pp. 8549--8560, 2018.

\bibitem{blalock2021multiplying}
D.~Blalock and J.~Guttag, ``Multiplying matrices without multiplying,''
  \emph{arXiv preprint arXiv:2106.10860}, 2021.

\bibitem{2021channel}
Y.~Wang, H.~Lu, and H.~Sun, ``Channel estimation in irs-enhanced mmwave system
  with super-resolution network,'' \emph{IEEE Communications Letters}, 2021.

\bibitem{ershadh2021computationally}
M.~Ershadh and M.~Meenakshi, ``A computationally lightest and robust neural
  network receiver for ultra wideband time hopping communication systems,''
  \emph{IEEE Transactions on Vehicular Technology}, vol.~70, no.~5, pp.
  4657--4668, 2021.

\bibitem{jiang2021ai}
P.~Jiang, T.~Wang, B.~Han, X.~Gao, J.~Zhang, C.-K. Wen, S.~Jin, and G.~Y. Li,
  ``Ai-aided online adaptive ofdm receiver: Design and experimental results,''
  \emph{IEEE Transactions on Wireless Communications}, 2021.

\bibitem{huber1965robust}
P.~J. Huber, ``A robust version of the probability ratio test,'' \emph{The
  Annals of Mathematical Statistics}, pp. 1753--1758, 1965.

\bibitem{dahlman20205g}
E.~Dahlman, S.~Parkvall, and J.~Skold, \emph{5G NR: The next generation
  wireless access technology}.\hskip 1em plus 0.5em minus 0.4em\relax Academic
  Press, 2020.

\bibitem{patzold2009two}
M.~Patzold, C.-X. Wang, and B.~O. Hogstad, ``Two new sum-of-sinusoids-based
  methods for the efficient generation of multiple uncorrelated rayleigh fading
  waveforms,'' \emph{IEEE Transactions on Wireless Communications}, vol.~8,
  no.~6, pp. 3122--3131, 2009.

\bibitem{press1989numerical}
W.~H. Press, W.~H. Press, B.~P. Flannery, S.~A. Teukolsky, W.~T. Vetterling,
  B.~P. Flannery, and W.~T. Vetterling, \emph{Numerical recipes in Pascal: the
  art of scientific computing}.\hskip 1em plus 0.5em minus 0.4em\relax
  Cambridge university press, 1989, vol.~1.

\bibitem{liu2014channel}
Y.~Liu, Z.~Tan, H.~Hu, L.~J. Cimini, and G.~Y. Li, ``Channel estimation for
  ofdm,'' \emph{IEEE Communications Surveys \& Tutorials}, vol.~16, no.~4, pp.
  1891--1908, 2014.

\bibitem{omar2008performance}
S.~Omar, A.~Ancora, and D.~T. Slock, ``Performance analysis of general
  pilot-aided linear channel estimation in lte ofdma systems with application
  to simplified mmse schemes,'' in \emph{2008 IEEE 19th International symposium
  on personal, indoor and mobile radio communications}.\hskip 1em plus 0.5em
  minus 0.4em\relax IEEE, 2008, pp. 1--6.

\bibitem{tang2007pilot}
Z.~Tang, R.~C. Cannizzaro, G.~Leus, and P.~Banelli, ``Pilot-assisted
  time-varying channel estimation for ofdm systems,'' \emph{IEEE Transactions
  on Signal Processing}, vol.~55, no.~5, pp. 2226--2238, 2007.

\bibitem{nissel2018doubly}
R.~Nissel, F.~Ademaj, and M.~Rupp, ``Doubly-selective channel estimation in
  fbmc-oqam and ofdm systems,'' in \emph{2018 IEEE 88th Vehicular Technology
  Conference (VTC-Fall)}.\hskip 1em plus 0.5em minus 0.4em\relax IEEE, 2018,
  pp. 1--5.

\bibitem{giannakis1998basis}
G.~B. Giannakis and C.~Tepedelenlioglu, ``Basis expansion models and diversity
  techniques for blind identification and equalization of time-varying
  channels,'' \emph{Proceedings of the IEEE}, vol.~86, no.~10, pp. 1969--1986,
  1998.

\bibitem{borah1999frequency}
D.~K. Borah and B.~Hart, ``Frequency-selective fading channel estimation with a
  polynomial time-varying channel model,'' \emph{IEEE Transactions on
  Communications}, vol.~47, no.~6, pp. 862--873, 1999.

\bibitem{ba2016layer}
J.~L. Ba, J.~R. Kiros, and G.~E. Hinton, ``Layer normalization,'' \emph{arXiv
  preprint arXiv:1607.06450}, 2016.

\bibitem{hendrycks2016gaussian}
D.~Hendrycks and K.~Gimpel, ``Gaussian error linear units (gelus),''
  \emph{arXiv preprint arXiv:1606.08415}, 2016.

\bibitem{zhang2017defense}
C.-L. Zhang, J.-H. Luo, X.-S. Wei, and J.~Wu, ``In defense of fully connected
  layers in visual representation transfer,'' in \emph{Pacific Rim Conference
  on Multimedia}.\hskip 1em plus 0.5em minus 0.4em\relax Springer, 2017, pp.
  807--817.

\bibitem{jakes1994microwave}
W.~C. Jakes and D.~C. Cox, \emph{Microwave mobile communications}.\hskip 1em
  plus 0.5em minus 0.4em\relax Wiley-IEEE press, 1994.

\bibitem{rodgers1985improvements}
D.~P. Rodgers, ``Improvements in multiprocessor system design,'' \emph{ACM
  SIGARCH Computer Architecture News}, vol.~13, no.~3, pp. 225--231, 1985.

\bibitem{liu2017learning}
Z.~Liu, J.~Li, Z.~Shen, G.~Huang, S.~Yan, and C.~Zhang, ``Learning efficient
  convolutional networks through network slimming,'' in \emph{Proceedings of
  the IEEE international conference on computer vision}, 2017, pp. 2736--2744.

\bibitem{takeda2019mimo}
M.~Y. Takeda, A.~Klautau, A.~Mezghani, and R.~W. Heath, ``{MIMO} channel
  estimation with non-ideal adcs: Deep learning versus gamp,'' in \emph{2019
  IEEE 29th International Workshop on Machine Learning for Signal Processing
  (MLSP)}.\hskip 1em plus 0.5em minus 0.4em\relax IEEE, 2019, pp. 1--6.

\bibitem{van2020optcomnet}
M.~van Lier, A.~Balatsoukas-Stimming, H.~Corporaal, and Z.~Zivkovic,
  ``Optcomnet: Optimized neural networks for low-complexity channel
  estimation,'' in \emph{ICC 2020-2020 IEEE International Conference on
  Communications (ICC)}.\hskip 1em plus 0.5em minus 0.4em\relax IEEE, 2020, pp.
  1--6.

\bibitem{han2015learning}
S.~Han, J.~Pool, J.~Tran, and W.~J. Dally, ``Learning both weights and
  connections for efficient neural networks,'' \emph{arXiv preprint
  arXiv:1506.02626}, 2015.

\bibitem{faust2021mechanisms}
T.~E. Faust, G.~Gunner, and D.~P. Schafer, ``Mechanisms governing
  activity-dependent synaptic pruning in the developing mammalian cns,''
  \emph{Nature Reviews Neuroscience}, vol.~22, no.~11, pp. 657--673, 2021.

\bibitem{hooker2019compressed}
S.~Hooker, A.~Courville, G.~Clark, Y.~Dauphin, and A.~Frome, ``What do
  compressed deep neural networks forget?'' \emph{arXiv preprint
  arXiv:1911.05248}, 2019.

\bibitem{narengerile2021deep}
N.~Narengerile, J.~Thompson, P.~Patras, and T.~Ratnarajah, ``Deep reinforcement
  learning-based beam training for spatially consistent millimeter wave
  channels,'' in \emph{2021 IEEE 32nd Annual International Symposium on
  Personal, Indoor and Mobile Radio Communications (PIMRC)}.\hskip 1em plus
  0.5em minus 0.4em\relax IEEE, 2021, pp. 579--584.

\end{thebibliography}

\end{document}